\documentclass[a4paper]{article}

\usepackage{authblk}
\usepackage{fullpage}
\usepackage[utf8]{inputenc} 
\usepackage{amsmath, amsthm, amssymb}
\usepackage[english]{babel}
\usepackage{pdfpages}
\usepackage{times}
\usepackage{paralist}
\usepackage{graphicx}
\usepackage{longtable}
\usepackage{booktabs}
\usepackage{tabu}
\usepackage[referable]{threeparttablex}
\usepackage{varwidth}
\usepackage{multirow}
\usepackage{xfrac}
\usepackage{hyperref}

\usepackage{subfig}

\usepackage{tikz}
\usepackage{comment}
\usepackage{parskip}
\usepackage{pbox}

\usepackage{todonotes}
\usepackage[ruled, linesnumbered]{algorithm2e}
\usepackage{tikz}
\usetikzlibrary{arrows,automata,shapes,decorations,decorations.markings,calc, matrix,decorations.pathmorphing}

\usepackage{enumerate,paralist}

\newtheorem{definition}{Definition}
\newtheorem{theorem}{Theorem}
\newtheorem{corollary}{Corollary}
\newtheorem{lemma}{Lemma}

\theoremstyle{remark}
\newtheorem{remark}{Remark}

\theoremstyle{definition}
\newtheorem{example}{Example}

\usepackage{graphicx} 
\newcommand{\Nats}{\mathbb{N}}
\newcommand{\Reals}{\mathbb{R}}

\newcommand{\Level}{\mathsf{Lv}}

\newcommand{\LD}{\mathsf{LD}}

\newcommand{\Distance}{d}

\newcommand{\Weight}{\mathsf{wt}}

\newcommand{\argmin}{\mathsf{argmin}}
\newcommand{\Bag}{B}
\newcommand{\Tree}{\mathrm{Tree}}

\newcommand{\Ushape}{\mathsf{U}}

\newcommand{\Preprocessalgo}{\mathsf{Preprocess}}
\newcommand{\Mergealgo}{\mathsf{Merge}}
\newcommand{\Updatealgo}{\mathsf{Update}}
\newcommand{\Queryalgo}{\mathsf{Query}}
\newcommand{\Dotvectoralgo}{\mathsf{dot\_vector}}
\newcommand{\Dotmatrixalgo}{\mathsf{dot\_matrix}}

\newcommand{\RMSDist}{\mathsf{RSMDistance}}
\newcommand{\Datastructure}{\mathcal{D}}

\newcommand{\Zero}{\overline{\mathbf{0}}}
\newcommand{\One}{\overline{\mathbf{1}}}
\newcommand{\True}{\mathsf{True}}
\newcommand{\False}{\mathsf{False}}

\newcommand{\Climbalgo}{\mathsf{Climb}}
\newcommand{\Up}{\mathsf{Up}}

\def\set#1{\{ #1 \}}

\makeatletter
\newcommand{\removelatexerror}{\let\@latex@error\@gobble}
\makeatother

\usepackage{thmtools}
\declaretheoremstyle[%
  qed=\qedsymbol%
]{mystyle} 
\declaretheorem[name={Proof},style=mystyle,unnumbered,
]{prf}

\title{Faster Algorithms for Algebraic Path Properties in RSMs with Constant Treewidth\thanks{This work has been supported by the Austrian Science Foundation (FWF)
under the NFN RiSE (S11405-07), FWF Grant P23499-N23, 
ERC Start grant (279307: Graph Games), and Microsoft faculty fellows award.}
}

\author[1]{Krishnendu Chatterjee\thanks{krish.chat@ist.ac.at}}
\author[1]{Rasmus Ibsen-Jensen\thanks{ribsen@ist.ac.at}}
\author[1]{Andreas Pavlogiannis\thanks{pavlogiannis@ist.ac.at}}
\author[2]{\\Prateesh Goyal\thanks{prateesh@iitb.ac.in}}

\affil[1]{IST Austria (Institute of Science and Technology Austria)}
\affil[2]{IIT Bombay (Indian Institute of Technology Bombay) Mumbai, India}

\begin{document}

%\authorinfo{}{}{}

\maketitle

\begin{abstract}
Interprocedural analysis is at the heart of numerous applications in 
programming languages, such as alias analysis, constant propagation, etc.
Recursive state machines (RSMs) are standard models for interprocedural 
analysis. 
We consider a general framework with RSMs where the transitions are labeled
from a semiring, and path properties are algebraic with semiring operations.
RSMs with algebraic path properties can model interprocedural 
dataflow analysis problems, the shortest path problem, the most probable path 
problem, etc.
The traditional algorithms for interprocedural analysis focus on path 
properties where the starting point is \emph{fixed} as the entry point of a 
specific method. 
In this work, we consider possible multiple queries as required in many 
applications such as in alias analysis.
The study of multiple queries allows us to bring in a very important 
algorithmic distinction between the resource usage of the \emph{one-time} 
preprocessing vs for \emph{each individual} query.
The second aspect that we consider is that the control flow graphs for most 
programs have constant treewidth.

Our main contributions are simple and implementable algorithms 
that support multiple queries for algebraic path properties for RSMs 
that have constant treewidth.
Our theoretical results show that our algorithms have small additional 
one-time preprocessing, but can answer subsequent queries 
significantly faster as compared to the current best-known solutions 
for several important problems, such as interprocedural reachability and 
shortest path.
We provide a prototype implementation for interprocedural reachability
and intraprocedural shortest path that gives a significant speed-up on 
several benchmarks.
\end{abstract}

%%\category{D.3.4}{Programming Languages}{Processors---Optimization}
%\category{F.3.2}{Logics and Meanings of Programs}{Semantics of Programming Languages---Program Analysis}

% general terms are not compulsory anymore, 
% you may leave them out
%%\terms Algorithms, Languages

%\keywords Interprocedural analysis, Constant treewidth graphs, Dataflow analysis, Reachability and shortest path.

\section{Introduction\label{sec:intro}}

\noindent{\em Interprocedural analysis and RSMs.}
Interprocedural analysis is one of the classic algorithmic problem  
in programming languages which is at the heart of numerous applications,
ranging from alias analysis, to data dependencies (modification and reference 
side effect), to constant propagation, to live and use analysis~\cite{Reps95, Sagiv96, Callahan86, Grove93, Land91, Knoop96, Cousot77, Giegerich81, Knoop92, Naeem08, Zhang14,Chatterjee15}.
%{\bf KRISH: More citations: say 5,14, 24, 22, 10, 13, 21 from IFDS.}
In seminal works~\cite{Reps95,Sagiv96} it was shown that a large class of 
interprocedural dataflow analysis problems can be solved in polynomial time. 
A standard model for interprocedural analysis is 
\emph{recursive state machines (RSMs)}~\cite{ABEGRY05} (aka \emph{supergraph} in~\cite{Reps95}).
A RSM is a formal model for control flow graphs of programs with recursion. 
We consider RSMs that consist of component state machines (CSMs), one for each method 
that has a unique entry and unique exit, and each CSM contains boxes which are labeled 
as CSMs that allows calls to other methods. 

\smallskip\noindent{\em Algebraic path properties.}
To specify properties of traces of a RSM we consider a very general framework,
where edges of the RSM are labeled from a partially complete semiring 
(which subsumes bounded and finite distributive semirings), and we refer 
to the labels of the edges as weights. 
For a given path, the weight of the path is the semiring product of the weights
on the edges of the path, and to choose among different paths we use the 
semiring plus operator. 
For example, (i)~with Boolean semiring (with semiring product as AND, and semiring
plus as OR) we can express the reachability property; 
(ii)~with tropical semiring (with real-edge weights, semiring product as standard sum, 
and semiring plus as minimum) we can express the shortest path property; and 
(iii)~with Viterbi semiring (with probability value on edges, semiring product
as standard multiplication and semiring plus as maximum) we can express the 
most probable path property.
The algebraic path properties expressed in our framework subsumes the 
IFDS/IDE frameworks~\cite{Reps95,Sagiv96} which consider finite semirings and meet 
over all paths as the semiring plus operator.
Since IFDS/IDE are subsumed in our framework, the large and important class of 
dataflow analysis problems that can be expressed in IFDS/IDE frameworks can also be 
expressed in our framework.

\smallskip\noindent{\em Two important aspects.}
In the traditional algorithms for interprocedural analysis, the starting point is 
typically \emph{fixed} as the entry point of a specific method. 
In graph theoretic parlance, graph algorithms can consider two types of queries: 
(i)~a \emph{pair query} that given nodes $u$ and $v$ (called $(u,v)$-pair query) asks 
for the algebraic path property from $u$ to $v$; and (ii)~a \emph{single-source} query 
that given a node $u$ asks for the answer of $(u,v)$-pair queries for all nodes $v$.
Thus the traditional algorithms for interprocedural analysis has focused on the 
answer for \emph{one} single-source query.
Moreover, the existing algorithms also consider that the input control flow graph 
is arbitrary, and do not exploit the fact that most control flow graphs satisfy 
several elegant structural properties.
In this work, we consider two new aspects, namely, (i)~\emph{multiple} pair and 
single-source queries, and  (ii)~exploit the fact that typically the control flow 
graphs of programs satisfy an important structural property called the 
\emph{constant treewidth property}.
We describe in details the two aspects.
\begin{compactitem}
\item \emph{Multiple queries.} We first describe the relevance of pair and multiple pair 
queries, and then the significance of even multiple single-source queries.
In alias analysis, the question is whether two pointers may point to the same object,
which is by definition modeled as a question between a pair of nodes.
Similarly, e.g., in constant propagation, given a function call, 
a relevant question is whether some variable remains constant within the entry and exit
of the function (in general it can be between a pair of nodes of the program).
This shows that the pair query problem, and the multiple pair queries are relevant 
in many applications.
Finally, consider a run-time optimization scenario, where the goal is to decide whether 
a variable remains constant from \emph{now on}, and this corresponds to a 
single-source query, where the starting point is the current execution point of 
the program.
Thus multiple pair queries and multiple single-source queries are relevant 
for several important static analysis problems.

\item \emph{Constant treewidth.}
A very well-known concept in graph theory is the notion of {\em treewidth} of 
a graph, which is a measure of how similar a graph is to a tree 
(a graph has treewidth~1 precisely if it is a tree)~\cite{Robertson84}. 
The treewidth of a graph is defined based on a {\em tree decomposition} of 
the graph~\cite{Halin76}, see Section~\ref{sec:definitions} for a formal definition. 
Beyond the mathematical elegance of the treewidth property for graphs,
there are many classes of graphs which arise in practice and have constant treewidth. 
The most important example is that the control flow graph for goto-free programs 
for many programming languages are of constant treewidth~\cite{Thorup98},
and it was also shown in~\cite{Gustedt02} that typically all Java programs have 
constant treewidth. 
An important property of constant-treewidth graphs is that the number of
edges is at most a constant factor larger than the number of nodes.
This has been considered in the comparison Tables~\ref{tab:IDFS-comparison} and \ref{tab:shortest_path-comparison}.

\end{compactitem}

\begin{table*}
\scriptsize
\renewcommand{\arraystretch}{1.2}
\centering
\begin{tabular}{|c|c|c|c|c|c|}
\hline &Preprocessing time & Space & Single source query & Pair query & Reference \\
\hline
Our & $O(\log n \cdot (n+h\cdot b))$ & $O(n\cdot \log n)$ & $O(n)$ & $O(1)$ &  Theorem~\ref{thm:rmsdist}\\
\cline{2-6}
Results & $O(\log n \cdot (n+h\cdot b))$ & $O(n)$ & $O(n)$ & $O(\log n)$ & Theorem~\ref{thm:rmsdist}\\
\hline
\end{tabular}
\caption{Interprocedural same-context algebraic path problem on RSMs with $b$ boxes and constant treewidth, for stack height $h$. \label{tab:main_result}}
\end{table*}

\begin{table*}
\scriptsize
\renewcommand{\arraystretch}{1.2}
\centering
\begin{tabular}{|c|c|c|c|c|c|}
\hline &Preprocessing time &  Space & Single source query & Pair query & Reference \\
\hline
\pbox[t]{20cm}{IDE/IFDS \\  (complete preprocessing)} & $O(n^2 \cdot |D|^3)$ & $O(n^2 \cdot |D|)$ & $O(n\cdot |D|)$ & $O(|D|)$ & \cite{Reps95,Sagiv96} \\
\hline
\pbox[t]{20cm}{IDE/IFDS \\  (no preprocessing)}  & - & $O(n \cdot |D|)$ & $O(n\cdot |D|^3)$ & $O(n\cdot |D|^3)$ & \cite{Reps95,Sagiv96}\\
\hline
Our & $O(|D|^2\cdot \log n\cdot (n+b\cdot |D|))$ & $O(n\cdot\log n\cdot |D|^2)$ & $O(n\cdot |D|^2)$ & $O( |D|^2)$ & Corollary~\ref{cor:ifds} \\
\cline{2-6}
Results & $O(n\cdot |D|^2+ \log n \cdot (b \cdot |D|^3 + n))$ & $O(n\cdot |D|^2)$ & $O(n\cdot |D|^2)$ & $O(\log n \cdot |D|^2)$ & Corollary~\ref{cor:ifds}\\
\hline
\end{tabular}
\caption{Interprocedural same-context algebraic path problem on RSMs with $b$ boxes and constant treewidth, where the semiring is over the subset of $|D|$ elements and 
the plus operator is the meet operator of the IFDS framework.
The special case of reachability is obtained when $|D|=1$. \label{tab:IDFS-comparison}}
\end{table*}

\begin{comment}
\begin{table*}
\small
\renewcommand{\arraystretch}{1.2}
\centering
\begin{tabular}{|c|c|c|c|c|c|}
\hline &Preprocessing time &  Single-source query & Pair query & Space & Reference \\
\hline
\pbox[t]{20cm}{IDE/IFDS \\  (complete preprocessing)} & $O(n^2)$ & $O(n)$ & $O(1)$ & $O(n^2)$ & \cite{Reps95,Sagiv96}\\
\hline
\pbox[t]{20cm}{IDE/IFDS \\  (no preprocessing)} & - & $O(n)$ & $O(n)$ & $O(n)$ & \cite{Reps95,Sagiv96}\\
\hline
Our & $O(n\cdot\log n)$ & $O(n)$ & $O(1)$ & $O(n\cdot\log n)$ & Corollary~\ref{cor:ifds} for $|D|=1$\\
\cline{2-6}
Results & $O(n\cdot \log n)$ & $O(n)$ & $O(\log n)$ & $O(n)$ & Corollary~\ref{cor:ifds} for $|D|=1$\\
\hline
\end{tabular}
\caption{Interprocedural same-context reachability for RSMs with constant treewidth. \label{tab:IDFS-comparison-reach}}
\end{table*}
\end{comment}

\begin{table*}
\scriptsize
\renewcommand{\arraystretch}{1.2}
\centering
\begin{tabular}{|c|c|c|c|c|c|}
\hline &Preprocessing time & Space &  Single-source query & Pair query & Reference \\
\hline
GPR (complete preprocessing) & $O(n^5)$ & $O(n^2)$ & $O(n)$ & $O(1)$ & \cite{Reps05,Reps07} \\
\hline
GPR (no preprocessing) & - & $O(n)$ &  $O(n^4)$ & $O(n^4)$ & \cite{Reps05,Reps07}\\
\hline
Our & $O(n^2\cdot\log n)$ & $O(n\cdot\log n)$ & $O(n)$ & $O(1)$ & Corollary~\ref{cor:shortest_paths} \\
\cline{2-6}
Results & $O(n^2\cdot \log n)$ & $O(n)$ & $O(n)$ & $O(\log n)$ & Corollary~\ref{cor:shortest_paths}\\
\hline
\end{tabular}
\caption{Interprocedural same-context shortest path for RSMs with constant treewidth. \label{tab:shortest_path-comparison}}
\end{table*}

\smallskip\noindent{\em Our contributions.}
In this work we consider RSMs where every CSM has constant treewidth, and the 
algorithmic question of answering multiple single-source and multiple pair 
queries, where each query is a \emph{same-context} query (a same-context query starts 
and ends with an empty stack, see~\cite{Chaudhuri95} for the significance of same-context queries).
In the analysis of multiple queries, there is a very important algorithmic 
distinction between \emph{one-time} preprocessing (denoted as the 
preprocessing time), and the work done for each individual query 
(denoted as the query time).
There are two end-points in the spectrum of tradeoff between the preprocessing and query 
resources that can be obtained by using the classical algorithms for one single-source query, 
namely, (i)~the \emph{complete preprocessing}, and (ii)~the \emph{no preprocessing}. 
In complete preprocessing, the single-source answer is precomputed with every node as the starting 
point (for example, in graph reachability this corresponds to computing the all-pairs reachability 
problem with the classical BFS/DFS algorithm~\cite{Cormen01}, or with fast matrix multiplication~\cite{Fischer71}).
In no preprocessing, there is no preprocessing done, and the algorithm 
for one single-source query is used on demand for each individual query. 
We consider various other possible tradeoffs in preprocessing vs query time.
Our main contributions are as follows:

\begin{compactenum}

\item \emph{(General result).}
Since we consider arbitrary semirings (i.e., not restricted to finite semirings)
we consider the stack height bounded problem, where the height of the stack is bounded
by a parameter $h$. 
While in general for arbitrary semirings there does not exist a bound on the stack 
height, if the semiring contains subsets of a finite universe $D$, and 
the semiring plus operator is intersection or union, then solving the problem 
with sufficiently large bound on the stack height is equivalent to solving the problem without any 
restriction on stack height.
Our main result is an algorithm where the one-time preprocessing phase requires 
$O(n\cdot \log n+h\cdot b \cdot \log n)$ semiring operations, and then each 
subsequent bounded stack height pair query can be answered in constant number 
of semiring operations, where 
$n$ is the number of nodes of the RSM and $b$ the number of boxes
(see Table~\ref{tab:main_result} and Theorem~\ref{thm:rmsdist}).
If we specialize our result to the IFDS/IDE setting with finite semirings from a finite universe 
of distributive functions $2^D\rightarrow 2^D$,
and meet over all paths as the semiring plus operator, then we obtain the 
results shown in Table~\ref{tab:IDFS-comparison} (Corollary~\ref{cor:ifds}).
For example, our approach with a factor of $O(\log n)$ overhead for one-time 
preprocessing, as compared no preprocessing, 
can answer subsequent pair queries by a factor of $O(n \cdot |D|)$ faster.
An important feature of our algorithms is that they are simple and implementable.

\item \emph{(Reachability and shortest path).}
We now discuss the significance of our result for the very important special 
cases of reachability and shortest path.
\begin{compactitem}
\item \emph{(Reachability).}
The result for reachability with full preprocessing, no preprocessing, 
and the various tradeoff that can be obtained by our approach is obtained from 
Table~\ref{tab:IDFS-comparison} by $|D|=1$.
For example for pair queries, full preprocessing requires quadratic time and 
space (for all-pairs reachability computation) and answers individual queries 
in constant time;
no preprocessing requires linear time and space for individual queries; 
whereas with our approach 
(i)~with almost-linear ($O(n \cdot \log n)$) preprocessing time and space 
we can answer individual queries in constant time, which is a 
significant (from quadratic to almost-linear) improvement over full preprocessing; 
or (ii)~with linear space and almost-linear preprocessing time we can answer queries 
in logarithmic time, which is a huge (from linear to logarithmic) improvement 
over no preprocessing.
For example, if we consider $O(n)$ pair queries, then both full preprocessing 
and no preprocessing in total require quadratic time, whereas 
our approach in total requires $O(n \cdot \log n + n \cdot \log n)= 
O(n \cdot \log n)$ time.
%Our results are presented in Corollary~\ref{cor:ifds} for $|D|=1$.

\item \emph{(Shortest path).}
We now consider the problem of shortest path, where the current best-known 
algorithm is for pushdown graphs~\cite{Reps05,Reps07} and we are not aware of any better 
bounds for RSMs (that have unique entries and exits).
The algorithm of~\cite{Reps05} is a polynomial-time algorithm of degree four, and 
the full preprocessing requires $O(n^5)$ time and quadratic space, and 
can answer single-source (resp. pair) queries in linear (resp. constant time);
whereas the no preprocessing requires $O(n^4)$ time and linear space for 
both single-source and pair queries.
In contrast, we show that 
(i)~with almost-quadratic ($O(n^2 \cdot \log n)$) preprocessing time and 
almost-linear space, we can answer single-source (resp. pair) queries in 
linear (resp. constant) time; or
(i)~with almost-quadratic preprocessing and linear space, 
we can answer single-source (resp. pair) queries in linear (resp. logarithmic) 
time. 
%%In case that witness paths are required, the cost increases to $O(|P|)$ per
%%node $v$, where $P$ is the witness path for node $v$.
%While the results of~\cite{} also reports a witness path in the given running 
%time, we can report a witness path with an additional factor of $O(k)$ in the 
%query times, where $k$ is the length of the path and is bounded by $n$.
Thus our approach provides a significant theoretical improvement over the 
existing approaches.
\end{compactitem}
There are two facts that are responsible for our improvement, the first is 
that we consider that each CSM of the RSM has constant treewidth, and 
the second is the tradeoff of one-time preprocessing and individual queries.
Also note that our results apply only to same-context queries.

\item \emph{(Experimental results).}
Besides the theoretical improvements, we demonstrate the effectiveness of our 
approach on several well-known benchmarks from programming languages.
We use the tool for computing tree decompositions from~\cite{Dijk06}, 
and all benchmarks of our experimental results have small treewidth.
We have implemented our algorithms for reachability (both intraprocedural and 
interprocedural) and shortest paths (only intraprocedural), and 
compare their performance against complete and no preprocessing approaches 
for same-context queries.
Our experimental results show that our approach obtains a significant 
improvement over the existing approaches (of complete and no preprocessing).

\end{compactenum}

\smallskip\noindent{\em Technical contribution.} 
Our main technical contribution is a dynamic algorithm 
(also referred to as incremental algorithm in graph algorithm literature) 
that given a graph with constant treewidth, after a preprocessing phase of 
$O(n\cdot \log n)$ semiring operations supports (1)~changing the label 
of an edge with $O(\log n)$ semiring operations; and (2)~answering pair queries with 
$O(\log n)$ semiring operations; and (3)~answering single-source queries with $O(n)$ 
semiring operations.
These results are presented in Theorem~\ref{thm:dynamization}.

\smallskip\noindent{\em Nice byproduct.}
Several previous works such as~\cite{Horwitz95} have stated the importance and asked for the 
development of data structures and analysis techniques to support dynamic 
updates. 
Though our main results are for the problem where the RSM is given and fixed,
our main technical contribution is a dynamic algorithm that can also be used
in other applications to support dynamic updates, and is thus also of 
independent interest.

\subsection{Related Work}
In this section we compare our work with several related work 
from interprocedural analysis as well as for constant treewidth
property.

\smallskip\noindent{\em Interprocedural analysis.}
Interprocedural analysis is a classic algorithmic problem in 
static analysis and several diverse applications have been studied
in the literature~\cite{Reps95, Sagiv96, Callahan86, Grove93, Land91, Knoop96, Cousot77, Giegerich81, Knoop92}.
%{\bf KRISH MANY CITES.}
Our work is most closely related to the IFDS/IDE frameworks introduced
in seminal works~\cite{Reps95,Sagiv96}.
In both IFDS/IDE framework the semiring is finite, and they study 
the algorithmic question of solving one single-source query.
While in our framework the semiring is not necessarily finite, we 
consider the stack height bounded problem.
We also consider the multiple pair and single-source, same-context queries, and 
the additional restriction that RSMs have constant treewidth. 
Our general result specialized to finite semirings (where the stack
height bounded problem coincides with the general problem) improves
the existing best known algorithms for the IFDS/IDE framework where
the RSMs have constant treewidth.
For example, the shortest path problem cannot be expressed in the IFDS/IDE
framework~\cite{Reps05}, but can be expressed in the GPR framework~\cite{Reps05,Reps07}.
The GPR framework considers the more general problem of weighted 
pushdown graphs, whereas we show that with the restriction to constant treewidth 
RSMs the bounds for the best-known algorithm can be significantly improved.
Finally, several works such as~\cite{Horwitz95} ask for on-demand interprocedural
analysis and algorithms to support dynamic updates, and our main technical contributions 
are algorithms to support dynamic updates in interprocedural analysis.

\smallskip\noindent{\em Recursive state machines (RSMs).}
Recursive state machines, which in general are equivalent to pushdown graphs,
have been studied as a formal model for interprocedural analysis~\cite{ABEGRY05}.
However, in comparison to pushdown graphs, RSMs are a more convenient 
formalism for interprocedural analysis.
Games on recursive state machines with modular strategies have been considered 
in~\cite{ATM06,CV12}, and subcubic algorithm for general RSMs with reachability has been 
shown in~\cite{Chaudhuri08}.
We focus on RSMs with unique entries and exits and with the restriction that the
components have constant tree width. RSMs with unique entries and exits
are less expressive than pushdown graphs, but remain a very natural model
for efficient interprocedural analysis~\cite{Reps95,Sagiv96}.

\smallskip\noindent{\em Treewidth of graphs.}
The notion of treewidth for graphs as an elegant mathematical tool
to analyze graphs was introduced in~\cite{Robertson84}.
The significance of constant treewidth in graph theory is huge 
mainly because several problems on graphs become complexity-wise easier.
Given a tree decomposition of a graph with low treewidth $t$, 
many NP-complete problems for arbitrary graphs can be solved in time polynomial 
in the size of the graph, but exponential in $t$~\cite{Arnborg89,Bern1987216,Bodlaender88,Bodlaender93,Bodlaender05}. 
Even for problems that can be solved in polynomial time, 
faster algorithms can be obtained for low treewidth graphs,
for example, for the distance problem~\cite{Chaudhuri95}.
The constant-treewidth property of graphs has also been used in the context
of logic: Monadic Second Order (MSO) logic is a very expressive logic, and a 
celebrated result of~\cite{Courcelle91} showed that for constant-treewidth graphs the 
decision questions for MSO can be solved in polynomial time; and 
the result of~\cite{Elberfeld10} shows that this can even be achieved in deterministic
log-space.
Dynamic algorithms for the special case of 2-treewidth graphs has been 
considered in~\cite{Bodlaender94} and extended to various tradeoffs by~\cite{Hagerup00};
and~\cite{L13} shows how to maintain the strongly connected component decomposition
under edge deletions for constant treewidth graphs.
However, none of these works consider RSMs or interprocedural analysis.
Various other models (such as probabilistic models of Markov decision processes
and games played on graphs for synthesis) with the constant-treewidth restriction 
have also been considered~\cite{CL13,Obdrzalek03}.
The problem of computing a balanced tree decomposition for a constant treewidth 
graph was considered in~\cite{Reed92}, and we use this algorithm in our 
preprocessing phase. 
More importantly, in the context of programming languages, it was shown by~\cite{Thorup98}
that the control flow graph for goto-free programs for many programming languages have 
constant treewidth.
This theoretical result was subsequently followed up in several practical approaches,
and it was shown in~\cite{Gustedt02} that though in theory Java programs might not
have constant treewidth, in practice Java programs do have constant treewidth.
We also use the existing tree-decomposition tool developed by~\cite{Dijk06} in our experimental results.

\section{Definitions}\label{sec:definitions}
We will in this section give definitions related to semirings, graphs, and recursive state machines. 

\subsection{Semirings}

\begin{definition}[Semirings]
\normalfont
We consider partially complete {\em semirings} $(\Sigma, \oplus, \otimes, \Zero, \One)$ where $\Sigma$ is a countable set, 
$\oplus$ and $\otimes$ are binary operators on $\Sigma$, and $\Zero, \One\in \Sigma$, and the following properties hold:
\begin{compactenum}
\item $\oplus$ is associative, commutative, and $\Zero$ is the neutral element,
\item $\otimes$ is associative, and $\One$ is the neutral element,
\item $\otimes$ distributes over $\oplus$,
\item $\oplus$ is infinitely associative,
\item $\otimes$ infinitely distributes over $\oplus$,
\item $\Zero$ absorbs in multiplication, i.e., $\forall a\in \Sigma: a\otimes \Zero=\Zero$.
\end{compactenum}
Additionally, we consider that semirings are equipped with a \emph{closure} operator $^*$,
such that $\forall s\in \Sigma:~s^*=\One\oplus (s\otimes s^*) =\One\oplus (s^*\otimes s)$.
\end{definition}

\subsection{Graphs and tree decomposition}

\begin{definition}[Graphs and weighted paths]
\normalfont
Let $G=(V,E)$ be a finite directed graph where $V$ is a set of $n$ nodes and
$E\subseteq V\times V$ is an edge relation of $m$ edges, along with a weight
function $\Weight:E\rightarrow \Sigma$ that assigns to each edge of $G$ an element from $\Sigma$. 
A path $P:u\rightsquigarrow v$ is a sequence of edges $(e_1,\dots, e_k)$ and 
each $e_i=(x_i,y_i)$ is such that $x_1=u$, $y_k=v$, and for all $1 \leq i \leq k-1$
we have $y_i=x_{i+1}$. The length of $P$ is $k-1$.
A path $P$ is \emph{simple} if no node repeats in the path (i.e., it does not
contain a cycle). A single node is by itself a $0$-length path. 
Given a path $P=(e_1,\dots , e_k)$, the weight of $P$ is $\otimes(P)=\bigotimes(\Weight(e_1),\dots, \Weight(e_k))$ if 
$|P|\geq 1$ else $\otimes(P)=\One$. Given nodes $u,v\in V$, the distance $\Distance(u,v)$
is defined as $\Distance(u,v)=\bigoplus_{P:u\rightsquigarrow v}\otimes(P)$, and $\Distance(u,v)=\Zero$
if no such $P$ exists.
%Note that since $\Zero$ absorbs in multiplication, paths $u\rightsquigarrow v$ that contain a pair
%of nodes $(x, y)\not \in E$ do not contribute to the distance $\Distance(u,v)$ (unless $\Distance(u,v)=\Zero$).
\end{definition}

\smallskip
\begin{definition}[Tree decomposition and treewidth~\cite{Robertson84,Bodlaender93}]
\normalfont
Given a graph $G=(V,E)$, a {\em tree-decomposition} $\Tree(G)=(V_T, E_T)$ is a tree such that
the following conditions hold:
\begin{compactenum}
\item $V_T=\{\Bag_0,\dots, \Bag_{n'-1}:$ for all $0\leq i\leq n'-1$, $\Bag_i\subseteq V\}$ and $\bigcup_{\Bag_i\in V_T}\Bag_i=V$.
\item For all $(u,v)\in E$ there exists $\Bag_i\in V_T$ such that $u,v\in \Bag_i$.
\item For all $i,j,k$ such that there exist paths $\Bag_i\rightsquigarrow \Bag_k$ and $\Bag_k\rightsquigarrow \Bag_j$ in $\Tree(G)$, 
we have $\Bag_i\cap \Bag_j\subseteq \Bag_k$.
\end{compactenum}
The sets $\Bag_i$ which are nodes in $V_T$ are called bags.
The {\em width} of a tree-decomposition $\Tree(G)$ is the size of the largest bag minus~1
and the {\em treewidth} of $G$ is the width of a minimum-width tree decomposition of $G$.
It follows from the definition that if $G$ has constant treewidth, then $m=O(n)$.
\end{definition}

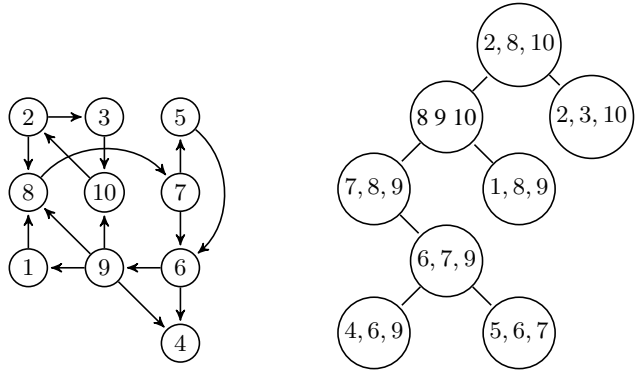
\begin{figure}
\newcommand{\distone}{5cm*0.2}
\centering
\begin{tikzpicture}[->,>=stealth',shorten >=1pt,auto,node distance=\distone,
                    semithick,scale=1 ]
      
\tikzstyle{every state}=[fill=white,draw=black,text=black,font=\small , inner sep=0.05cm, minimum size=0.5cm]
\tikzstyle{invis}=[fill=white,draw=white,text=white,font=\small , inner sep=-0.05cm]

\node[state] (v1) at (0,0) {$1$};
\node[state,above of=v1] (v8) {$8$};
\node[state,right of=v1] (v9) {$9$};
\node[state,above of=v8] (v2) {$2$};
\node[state,above of=v9] (v10) {$10$};
\node[state,above of=v10] (v3) {$3$};
\node[state,right of=v9] (v6) {$6$};
\node[state,below of=v6] (v4) {$4$};
\node[state,above of=v6] (v7) {$7$};
\node[state,above of=v7] (v5) {$5$};

\path (v1) edge (v8)
(v8) edge[bend left=50] (v7)
(v9) edge (v8) edge (v1) edge (v4) edge (v10)
(v2) edge (v3) edge (v8)
(v10) edge (v2)
(v3) edge (v10)
(v6) edge (v9) edge (v4)
(v7) edge (v5) edge (v6)
(v5) edge[bend left=50] (v6);

%\renewcommand{\distone}{5cm*0.38}
%\node[below of =v5] (ra) {\Large$\Rightarrow$};

\tikzstyle{every state}=[fill=white,draw=black,text=black,font=\small , inner sep=0.05cm, minimum size=0.9cm]
\renewcommand{\distone}{4.5cm*0.3}

%\node[state,below of=v9,node distance=sqrt(2)*\dist] (v8910) {8 9 10};
\node[state,] (v8910) at (5.5,2) {8 9 10};
\node[state,below right of=v8910] (v189) {$1,8,9$};
\node[state,above right of=v8910] (v2810) {$2,8,10$};
\node[state,below right of=v2810] (v2310) {$2,3,10$};
\node[state,below left of=v8910] (v789) {$7,8,9$};
\node[state,below right of=v789] (v679) {$6,7,9$};
\node[state,below left of=v679] (v469) {$4,6,9$};
\node[state,below right of=v679] (v567) {$5,6,7$};

\path[-] (v189) edge (v8910)
(v8910) edge (v2810) edge (v789)
(v2810) edge (v2310)
(v789) edge (v679)
(v679) edge (v469) edge (v567);

\end{tikzpicture}
\caption{A graph $G$ with treewidth $2$ (left) and a corresponding tree-decomposition $\Tree(G)$ (right).\label{fig:G}}
\end{figure}

\smallskip
\begin{example}[Graph and tree decomposition]
The treewidth of a graph $G$ is an intuitive measure which represents the 
proximity of $G$ to a tree, though $G$ itself not a tree.
The treewidth of $G$ is $1$ precisely if $G$ is itself a tree~\cite{Robertson84}.
Consider an example graph and its tree decomposition shown in Figure~\ref{fig:G}.
It is straightforward to verify that all the three conditions of 
tree decomposition are met.
Each node in the tree is a bag, and labeled by the set of nodes
it contains.
Since each bag contains at most three nodes, the tree decomposition by 
definition has treewidth~2.
\end{example}

\smallskip\noindent{\em Intuitive meaning of tree decomposition.}
In words, the tree-decomposition $\Tree(G)$ is a tree where every node 
(bag) is subset of nodes of $G$, such that:
(1)~every vertex in $G$ belongs to some bag; 
(2)~every edge in $G$ also belongs to some bag; and 
(3)~for every node $v$ of $G$, for every subpath in $\Tree(G)$, if $v$ 
appears in the endpoints of the path, then it must appear all along the path.

\smallskip\noindent{\bf Separator property.}
Given a graph $G$ and its tree decomposition $\Tree(G)$, 
note that for each bag $\Bag$ in $\Tree(G)$, if we remove the set of nodes
in the bag, then the graph splits into possibly multiple components 
(i.e., each bag is a separator for the graph).
In other words, every bag acts as a \emph{separator} of the graph.

\smallskip\noindent{\bf Notations for tree decomposition.}
Let $G$ be a graph,  $T=\Tree(G)$, and $\Bag_0$ be the root of $T$. 
Denote with $\Level\left(\Bag_i\right)$ the depth of $\Bag_i$ in $T$, with 
$\Level\left(\Bag_0\right)=0$. 
For $u\in V$, we say that a bag $\Bag$ \emph{introduces} $u$ if $\Bag$ 
is the bag with the smallest level among all bags that contain $u$,
i.e., $\Bag_u=\arg\min_{\Bag\in V_T:~u\in \Bag}\Level\left(\Bag\right)$. 
By definition, there is exactly one bag introducing each node $u$. 
We often write $\Bag_u$ for the bag that introduces the node $u$, 
and denote with $\Level(u)=\Level\left(\Bag_u\right)$.  
Finally, we denote with $\Bag_{(u,v)}$ the bag of the 
highest level that introduces one of $u$, $v$.
A tree-decomposition $\Tree(G)$ is \emph{semi-nice} if $\Tree(G)$ is
a binary tree, and every bag introduces at most one node.

\smallskip
\begin{example}
In the example of Figure~\ref{fig:G}, the bag $\{2,8,10\}$ is the root
of $\Tree(G)$, the level of node $9$ is $\Level(9)=\Level(\{8,9,10\})=1$, and the 
bag of the edge $(9,1)$ is $\Bag_{(9,1)}=\{1,8,9\}$.
\end{example}

\smallskip
\begin{theorem}\label{thm:seminice}
(1)~For every graph there exists a semi-nice tree decomposition that achieves the treewidth of G and uses $n' = O(n)$ bags~\cite{Kloks94}.
(2)~For constant treewidth graphs, a balanced tree decomposition can be obtained in $O(n\cdot\log n)$
time (i.e., every simple path $\Bag_0\rightsquigarrow \Bag_i$ in $\Tree(G)$ has length $O(\log n)$)~\cite{Reed92}. 
\end{theorem}

\smallskip\noindent{\bf The algebraic path problem on graphs of constant treewidth.}
Given $G=(V,E)$, a balanced, semi-nice tree-decomposition $\Tree(G)$ of $G$ with constant treewidth $t=O(1)$,
a partially complete semiring $(\Sigma, \oplus, \otimes, \Zero, \One)$,
a weight function $\Weight:E\rightarrow \Sigma$, the algebraic path problem on input $u,v\in V$,
asks for the distance $\Distance(u,v)$ from node $u$ to node $v$. 
In addition, we allow the weight function to change between successive queries.
We measure the time complexity of our algorithms in number of operations,
with each operation being either a basic machine operation,
or an application of one of the operators of the semiring.
%We consider that one application of the $\oplus$ and $\otimes$ operator costs time $O(\Cost)$.

\subsection{Recursive state machines}

\begin{definition}[RSMs and CSMs]
\normalfont
A {\em single-entry single-exit recursive state machine} (RSM from now on) over an alphabet $\Sigma$, as defined in~\cite{ABEGRY05}, consists of a set $\set{A_1,A_2,\dots, A_k}$, such that for each $1\leq i\leq k$, the {\em component state machine} (CSM) $A_i=(B_i, Y_i,V_i, E_i,\Weight_i)$, where $V_i=N_i\cup  \{En_i\}\cup \{Ex_i\}\cup C_i\cup R_i$, consists of:
\begin{compactitem}
\item A set $B_i$ of {\em boxes}.
\item A map $Y_i$, mapping each box in $B_i$ to an index in $\set{1,2,\dots,k}$. We say that a box $b\in B_i$ {\em corresponds} to the CSM with index $Y_i(b)$.
\item A set $V_i$ of {\em nodes}, consisting of the union of the sets $N_i$, $\{En_i\}$, $\{Ex_i\}$, $C_i$ and $R_i$. The number $n_i$ is the size of $V_i$. Each of these sets, besides $V_i$, are w.l.o.g. assumed to be pairwise disjoint.
\begin{compactitem}
\item The set $N_i$ is the set of {\em internal nodes}.
\item The node $En_i$ is the {\em entry node}.
\item The node $Ex_i$ is the {\em exit node}.
\item The set $C_i$ is the set of {\em call nodes}. Each call node is a pair $(x,b)$, where $b$ is a box in $B_i$ and $x$ is the entry node $En_{Y_i(b)}$ of the corresponding CSM with index $Y_i(b)$.
\item The set $R_i$ is the set of {\em return nodes}. Each return node is a pair $(y,b)$, where $b$ is a box in $B_i$ and $y$ is the exit node $Ex_{Y_i(b)}$ of the corresponding CSM with index $Y_i(b)$. 
\end{compactitem}
\item A set $E_i$ of {\em internal edges}. Each edge is a pair in $(N_i\cup \{En_i\}\cup R_i)\times (N_i\cup \{Ex_i\}\cup C_i)$.
\item A map $\Weight_i$, mapping each edge in $E_i$ to a label in $\Sigma$.
\end{compactitem}
\end{definition}

\smallskip
\begin{definition}[Control flow graph of CSMs and treewidth of RSMs]\label{def:cfg}
\normalfont
Given a RSM $A=\set{A_1,A_2,\dots, A_k}$, the \emph{control flow graph} $G_i=(V_i,E_i')$ for CSM $A_i$ 
consists of $V_i$ as the set of vertices and $E_i'$ as the set of edges, where $E_i'$ consists of the 
edges $E_i$ of $A_i$ and for each box $b$, each call node $(v,b)$  of that box (i.e. for $v=En_{Y_i(b)}$) 
has an edge to each return node $(v',b)$ of that box (i.e. for $v'=Ex_{Y_i(b)}$).
We say that the RSM has {\em treewidth} $t$, if $t$ is the smallest integer such that for each index $1\leq i \leq k$, 
the graph $G_i=(V_i,E_i')$ has treewidth at most $t$. 
Programs are naturally represented as RSMs, where the control flow graph of each method of a program is represented as a CSM.
\end{definition}

\smallskip
\begin{example}[RSM and tree decomposition]
Figure~\ref{fig:rsm} shows an example of a program for matrix multiplication consisting of 
two methods (one for vector multiplication invoked by the one for matrix multiplication).
The corresponding control flow graphs, and their tree decompositions
that achieve treewidth $2$ are also shown in the figure. 
\end{example}

\smallskip\noindent{\bf Box sequences.}
For a sequence $L$ of boxes and a box $b$, we denote with $L\circ b$ the concatenation of $L$ and $b$. Also, $\emptyset$ is the empty sequence of boxes.

\smallskip\noindent{\bf Configurations and global edges.} A {\em configuration} of a RSM is a pair $(v,L)$, where $v$ is a node in $(N_i\cup \{En_i\}\cup R_i)$ and $L$ is a sequence of boxes. The {\em stack height} of a configuration $(v,L)$ is the number of boxes in the sequence $L$.
The set of {\em global edges} $E$ are edges between configurations. The map $\Weight$ maps each edge in $E$ to a label in $\Sigma$.
We have that there is an edge between configuration $c_1=(v_1,L_1)$, where $v_1\in V_i$, and configuration $c_2=(v_2,L_2)$ with label $\sigma=\Weight(c_1,c_2)$ if and only if one of the following holds:
\begin{compactitem}
\item{\bf Internal edge:} We have that $v_2$ is an {\em internal} node in $N_i$ and each of the following (i)~$L_1=L_2$; and (ii)~$(v_1,v_2)\in E_i$; and (iii)~$\sigma=\Weight_i((v_1,v_2))$. 
\item{\bf Entry edge:} We have that  $v_2$ is the {\em entry} node $En_{Y_i(b)}$, for some box $b$,
 and each of the following (i)~$L_1\circ b=L_2$; and (ii)~$(v_1,(v_2,b))\in E_i$; and (iii)~$\sigma=\Weight_i((v_1,(v_2,b)))$.
\item{\bf Return edge:} We have that $v_2=(v,b)$ is a {\em return} node, for some exit node $v=Ex_i$ and some box $b$ and each of the following (i)~$L_1=L_2\circ b$; and (ii)~$(v_1,v)\in E_i$; and (iii)~$\sigma=\Weight_i((v_1,v))$.
\end{compactitem}

Note that in a configuration $(v,L)$, the node $v$ cannot be $Ex_i$ or in $C_i$. In essence, the corresponding configuration is at the corresponding return node, instead of at the exit node, or corresponding entry node, instead of at the call node, respectively.

\smallskip\noindent{\bf Execution paths.} 
An {\em execution path} is a sequence of configurations and labels $P=\langle c_1,\sigma_1,c_2,\sigma_2\dots,\sigma_{\ell-1},c_\ell\rangle$, such that for each integer $i$ where $1\leq i\leq \ell-1$, we have that $(c_i,c_{i+1})\in E$ and $\sigma_i=\Weight(c_i,c_{i+1})$. We call $\ell$ the {\em length} of $P$. Also, we say that the stack height of a execution path is the maximum stack height of a configuration in the execution path.
For a pair of configurations $c,c'$, the set $c\rightsquigarrow c'$, is the set of execution paths $\langle c_1,\sigma_1,c_2,\sigma_2\dots,\sigma_{\ell-1},c_\ell\rangle$, for any $\ell$, where $c=c_1$ and $c'=c_\ell$. 
For a set $S$ of execution paths, the set $B(S,h)\subseteq S$ is the subset of execution paths, with stack height at most $h$.
Given a partially complete semiring $(\Sigma, \oplus, \otimes, \Zero, \One)$, the {\em distance} of a execution path $P=\langle c_1,\sigma_1,c_2,\sigma_2\dots,\sigma_{\ell-1},c_\ell\rangle$ is $\otimes(P)=\bigotimes(\sigma_1,\dots, \sigma_{\ell-1})$ (the empty product is $\One$). 
Given configurations $c,c'$, the {\em configuration distance} $\Distance(c,c')$
is defined as $\Distance(c,c')=\bigoplus_{P:c\rightsquigarrow c'}\otimes(P)$ (the empty sum is $\Zero$). Also, given configurations $c,c'$ and a stack height $h$, where $c'$ is $h$-reachable from $c$, the {\em bounded height configuration distance} $\Distance(c,c',h)$ is defined as $\Distance(c,c',h)=\bigoplus_{P:B(c\rightsquigarrow c',h)}\otimes(P)$. Note that the above definition of execution paths only allows for so called \emph{valid} paths~\cite{Reps95,Sagiv96}, i.e., paths that
fully respect the calling contexts of an execution.

\smallskip\noindent{\bf The algebraic path problem on RSMs of constant tree-width.}
Given (i)~a RSM $A=\set{A_1,A_2,\dots, A_k}$; and (ii)~for each $1\leq i \leq k$ a balanced, semi-nice tree-decomposition $\Tree(A_i):=\Tree((V_i,E_i'))$ 
with constant treewidth at most $t=O(1)$; and (iii)~a partially complete semiring $(\Sigma, \oplus, \otimes, \Zero, \One)$, 
the {\em algebraic path problem} on input nodes $u,v$,
asks for the distance $\Distance((u,\emptyset),(v,\emptyset))$, i.e. the distance between the configurations with the empty stack. 
Similarly, also given a height $h$, the {\em bounded height algebraic path problem} on input configurations $c,c'$,
asks for the distance $\Distance((u,\emptyset),(v,\emptyset),h)$. When it is clear from the context, we will write
$\Distance(u,v)$ to refer to the algebraic path problem of nodes $u$ and $v$ on RSMs.

\begin{remark}\label{rem:same_context}
Note that the empty stack restriction implies that $u$ and $v$ are nodes of 
the same CSM.
However, the paths from $u$ to $v$ are, in general, interprocedural, and 
thus involve invocations and returns from other CSMs. 
This formulation has been used before in terms of \emph{same-context}~\cite{Chaudhuri08}
and \emph{same-level}~\cite{Reps95} realizable paths and has several applications in program 
analysis, e.g. by capturing balanced parenthesis-like properties used in 
alias analysis~\cite{Sridharan05}.
\end{remark}

%{\bf TODO: Write that RSMs represent CFG of programs. Refer to Figure for example.}

\begin{figure*}[!htb]
\scalebox{0.9}{%
\renewcommand{\thealgocf}{}
\removelatexerror
\centering
\begin{tikzpicture}[thick, >=latex,
pre/.style={<-,shorten >= 1pt, shorten <=1pt, thick},
post/.style={->,shorten >= 1pt, shorten <=1pt,  thick},
und/.style={very thick, draw=gray},
bag/.style={ellipse, minimum height=7mm,minimum width=14mm,draw=gray!80, line width=1pt},
internal/.style={circle,draw=black!80, inner sep=2, minimum size=4.5mm},
entry/.style={isosceles triangle, shape border rotate=-90, isosceles triangle stretches, minimum width=8mm, minimum height=3.6mm, draw=black!80, inner sep=0},
exit/.style={isosceles triangle, shape border rotate=90, isosceles triangle stretches, minimum width=8mm, minimum height=3.6mm, draw=black!80, inner sep=0},
call/.style={isosceles triangle, shape border rotate=0, isosceles triangle stretches, minimum width=8mm, minimum height=3.6mm, draw=black!80, inner sep=0},
return/.style={isosceles triangle, shape border rotate=180, isosceles triangle stretches, minimum width=8mm, minimum height=3.6mm, draw=black!80, inner sep=0},
%exit/.style={circle,draw=black!80, inner sep=2, minimum size=4pt},
%call/.style={circle,draw=black!80, inner sep=2, minimum size=4pt},
%return/.style={circle,draw=black!80, inner sep=2, minimum size=4pt},
virt/.style={circle,draw=black!50,fill=black!20, opacity=0}]

\newcommand{\ynodestep}{-0.9}
\newcommand{\xnodestep}{0.6}
\newcommand{\xdisposition}{5}
\newcommand{\xcaptiondisposition}{1.3}
\newcommand{\xtextdisposition}{0.8}
\newcommand{\legendy}{2.5}
\node[internal] at (\xcaptiondisposition+-3, \legendy) {};
\node []	at (\xcaptiondisposition+-3+\xtextdisposition, \legendy) {internal};
\node[entry] at (\xcaptiondisposition+0, \legendy) {};
\node []	at (\xcaptiondisposition+0+\xtextdisposition, \legendy) {entry};
\node[exit] at (\xcaptiondisposition+3, \legendy) {};
\node []	at (\xcaptiondisposition+3+\xtextdisposition, \legendy) {exit};
\node[call] at (\xcaptiondisposition+6, \legendy) {};
\node []	at (\xcaptiondisposition+6+\xtextdisposition, \legendy) {call};
\node[return] at (\xcaptiondisposition+9, \legendy) {};
\node []	at (\xcaptiondisposition+9+\xtextdisposition, \legendy) {return};

\node	[text width = 7cm] at (0,0)	{%
\begin{varwidth}{\linewidth}

\begin{algorithm}[H]
\small
%\TitleOfAlgo{$\Dotvectoralgo$}
\small
\SetAlgoNoLine
\SetAlgorithmName{Method}{method}
\DontPrintSemicolon
%\setstretch{1.05}
\caption{$\Dotvectoralgo$}\label{algo:dotvector}
\KwIn{$x,y\in \Reals^n$}
\KwOut{The dot product $x^\top y$}
$\mathsf{result}\leftarrow 0$\\
\For{$i\leftarrow 1$ \KwTo $n$}{
$z\leftarrow  x[i]\cdot y[i]$\\
$\mathsf{result}\leftarrow \mathsf{result} +z$\\
}
\Return{$\mathsf{result}$}
\end{algorithm}

\end{varwidth}
};

\newcommand{\secondlineybias}{0.6}
\node	[text width = 7cm] at (0,-5+ \secondlineybias)	{%
\begin{varwidth}{\linewidth}

\begin{algorithm}[H]
\small
\SetAlgoNoLine
\SetAlgorithmName{Method}{Method}
\DontPrintSemicolon
%\setstretch{1.05}
\caption{$\Dotmatrixalgo$}\label{algo:dotmatrix}
\KwIn{$A\in \Reals^{n\times k}, B\in \Reals^{k\times m}$}
\KwOut{The dot product $A\times B$}
$C\leftarrow $ zero matrix of size $n\times m$\\
\For{$i\leftarrow 1$ \KwTo $n$}{
\For{$j\leftarrow 1$ \KwTo $m$}{
Call $\Dotvectoralgo(A[i,:], B[:,j])$\\
\label{line:call}
$C[i,j]\leftarrow $ the value returned by the call of line~\ref{line:call}\\
}
}
\Return{$C$}
\end{algorithm}
\setcounter{algocf}{0}

\end{varwidth}
};

\newcommand{\globalxdisposition}{6}
\newcommand{\globalydisposition}{1.5}
\newcommand{\ynodestep}{-0.9}
\newcommand{\xnodestep}{0.6}
\newcommand{\xdisposition}{3}

\newcommand{\ybagstep}{-1}
\newcommand{\xbagstep}{0.9}

%test
%\node[entry] at (\globalxdisposition+0,\globalydisposition+0*\ynodestep) {};

\node	[entry]		(x1)	at	(\globalxdisposition+0,\globalydisposition+0*\ynodestep)		{$1$};
\node	[internal]		(x2)	at	(\globalxdisposition+0,\globalydisposition+1*\ynodestep)		{$2$};
\node	[internal]		(x3)	at	(\globalxdisposition+-\xnodestep,\globalydisposition+2*\ynodestep)		{$3$};
\node	[internal]		(x4)	at	(\globalxdisposition+-\xnodestep,\globalydisposition+3*\ynodestep)		{$4$};
\node	[internal]		(x5)	at	(\globalxdisposition+\xnodestep,\globalydisposition+2*\ynodestep)		{$5$};
\node	[exit]		(x6)	at	(\globalxdisposition+\xnodestep,\globalydisposition+3*\ynodestep)		{$6$};

\draw [->, thick] (x1) to 	(x2);
\draw [->, thick] (x2) to 	(x3);
\draw [->, thick] (x2) to 	(x5);
\draw [->, thick] (x3) to 	(x4);
\draw [->, thick, bend right =20] (x4) to 	(x2);
\draw [->, thick] (x5) to 	(x6);

\newcommand{\ybagstep}{-1}
\newcommand{\xbagstep}{0.9}

\node	[bag]		(b1)	at	(\globalxdisposition+\xdisposition,\globalydisposition+0*\ybagstep)		{$1$};
\node	[bag]		(b2)	at	(\globalxdisposition+\xdisposition,\globalydisposition+1*\ybagstep)		{$1,2$};
\node	[bag]		(b3)	at	(\globalxdisposition+\xdisposition-\xbagstep,\globalydisposition+2*\ybagstep)		{$2,3$};
\node	[bag]		(b4)	at	(\globalxdisposition+\xdisposition-\xbagstep,\globalydisposition+3*\ybagstep)		{$2,3,4$};
\node	[bag]		(b5)	at	(\globalxdisposition+\xdisposition+\xbagstep,\globalydisposition+2*\ybagstep)		{$2,5$};
\node	[bag]		(b6)	at	(\globalxdisposition+\xdisposition+\xbagstep,\globalydisposition+3*\ybagstep)		{$5,6$};

\draw [-, thick] (b1) to 	(b2);
\draw [-, thick] (b2) to 	(b3);
\draw [-, thick] (b3) to 	(b4);
\draw [-, thick] (b2) to 	(b5);
\draw [-, thick] (b5) to 	(b6);

\newcommand{\baglabel}{1.1}
\node	at (\globalxdisposition+\xdisposition+\baglabel,\globalydisposition+0*\ybagstep)		{\large $\Bag_1$};
\node	at	(\globalxdisposition+\xdisposition+\baglabel,\globalydisposition+1*\ybagstep)		{\large $\Bag_2$};
\node	at	(\globalxdisposition+\xdisposition-\xbagstep-\baglabel,\globalydisposition+2*\ybagstep)		{\large $\Bag_3$};
\node	at	(\globalxdisposition+\xdisposition-\xbagstep-\baglabel,\globalydisposition+3*\ybagstep)		{\large $\Bag_4$};
\node	at	(\globalxdisposition+\xdisposition+\xbagstep+\baglabel,\globalydisposition+2*\ybagstep)		{\large $\Bag_5$};
\node	at	(\globalxdisposition+\xdisposition+\xbagstep+\baglabel,\globalydisposition+3*\ybagstep)		{\large $\Bag_6$};

\renewcommand{\globalydisposition}{-3.2}

\node	[entry]		(y1)	at	(\globalxdisposition+0,\globalydisposition+0*\ynodestep+ \secondlineybias)		{$1$};
\node	[internal]		(y2)	at	(\globalxdisposition+0,\globalydisposition+1*\ynodestep+ \secondlineybias)		{$2$};
\node	[internal]		(y3)	at	(\globalxdisposition+0,\globalydisposition+2*\ynodestep+ \secondlineybias)		{$3$};
\node	[call]		(y4)	at	(\globalxdisposition+0*\xnodestep,\globalydisposition+3*\ynodestep+ \secondlineybias)		{$4$};
\node	[return]		(y5)	at	(\globalxdisposition+0*\xnodestep,\globalydisposition+4*\ynodestep+ \secondlineybias)		{$5$};
\node	[internal]		(y6)	at	(\globalxdisposition+-\xnodestep,\globalydisposition+3*\ynodestep+ \secondlineybias)		{$6$};
\node	[internal]		(y7)	at	(\globalxdisposition+2*\xnodestep,\globalydisposition+2*\ynodestep+ \secondlineybias)		{$7$};
\node	[exit]		(y8)	at	(\globalxdisposition+2*\xnodestep,\globalydisposition+3*\ynodestep+ \secondlineybias)		{$8$};

\draw [->, thick] (y1) to 		(y2);
\draw [->, thick] (y2) to 		(y3);
\draw [->, thick] (y3) to 		(y4);
\draw [->, thick] (y3) to 		(y6);
\draw [->, thick] (y4) to 		(y5);
\draw [->, thick, bend right =40] (y5) to 	(y3);
\draw [->, thick, bend left = 20] (y6) to 	(y2);
\draw [->, thick] (y2) to 		(y7);
\draw [->, thick] (y7) to 		(y8);

\node	[bag]		(d1)	at	(\globalxdisposition+\xdisposition,\globalydisposition+0*\ybagstep+ \secondlineybias)		{$1$};
\node	[bag]		(d2)	at	(\globalxdisposition+\xdisposition,\globalydisposition+1*\ybagstep+ \secondlineybias)		{$1,2$};
\node	[bag]		(d3)	at	(\globalxdisposition+\xdisposition-\xbagstep,\globalydisposition+2*\ybagstep+ \secondlineybias)		{$2,3$};
\node	[bag]		(d4)	at	(\globalxdisposition+\xdisposition-2*\xbagstep,\globalydisposition+3*\ybagstep+ \secondlineybias)		{$3,4$};
\node	[bag]		(d5)	at	(\globalxdisposition+\xdisposition-2*\xbagstep,\globalydisposition+4*\ybagstep+ \secondlineybias)		{$3,4,5$};
\node	[bag]		(d6)	at	(\globalxdisposition+\xdisposition-0*\xbagstep,\globalydisposition+3*\ybagstep+ \secondlineybias)		{$2,3,6$};
\node	[bag]		(d7)	at	(\globalxdisposition+\xdisposition+\xbagstep,\globalydisposition+2*\ybagstep+ \secondlineybias)		{$2,7$};
\node	[bag]		(d8)	at	(\globalxdisposition+\xdisposition+2*\xbagstep,\globalydisposition+3*\ybagstep+ \secondlineybias)		{$7,8$};

\draw [-, thick] (d1) to 		(d2);
\draw [-, thick] (d2) to 		(d3);
\draw [-, thick] (d3) to 		(d4);
\draw [-, thick] (d4) to 		(d5);
\draw [-, thick] (d3) to 		(d6);
\draw [-, thick] (d2) to 		(d7);
\draw [-, thick] (d7) to 		(d8);

\node at	(\globalxdisposition+\xdisposition+\baglabel,\globalydisposition+0*\ybagstep+ \secondlineybias)		{\large $\Bag_1$};
\node	at	(\globalxdisposition+\xdisposition+\baglabel,\globalydisposition+1*\ybagstep+ \secondlineybias)		{\large $\Bag_2$};
\node	at	(\globalxdisposition+\xdisposition-\xbagstep-\baglabel,\globalydisposition+2*\ybagstep+ \secondlineybias)		{\large $\Bag_3$};
\node	at	(\globalxdisposition+\xdisposition-2*\xbagstep-\baglabel,\globalydisposition+3*\ybagstep+ \secondlineybias)		{\large $\Bag_4$};
\node	at	(\globalxdisposition+\xdisposition-2*\xbagstep-\baglabel,\globalydisposition+4*\ybagstep+ \secondlineybias)		{\large $\Bag_5$};
\node	at	(\globalxdisposition+\xdisposition-0*\xbagstep,\globalydisposition+3*\ybagstep+ \secondlineybias-\baglabel/1.5)		{\large $\Bag_6$};
\node	at	(\globalxdisposition+\xdisposition+\xbagstep+\baglabel,\globalydisposition+2*\ybagstep+ \secondlineybias)		{\large $\Bag_7$};
\node	at (\globalxdisposition+\xdisposition+2*\xbagstep,\globalydisposition+3*\ybagstep+ \secondlineybias-\baglabel/1.5)		{\large $\Bag_8$};

\end{tikzpicture}
}
\caption{Example of a program consisting of two methods, their control flow graphs $G_i=(V_i,E_i')$ where nodes correspond to line numbers,
 and the corresponding tree decompositions, each one achieving treewidth $2$.}\label{fig:rsm}
\end{figure*}

\subsection{Problems}
We note that a wide range of interprocedural problems can be formulated as bounded height algebraic path problems.
\begin{compactenum}
\item \emph{Reachability} i.e., given nodes $u$, $v$ in the same CSM, is there a path from $u$ to $v$?
The problem can be formulated on the boolean semiring $(\{\True, \False\}, \lor, \land,  \False, \True)$.
\item \emph{Shortest path} i.e., given a weight function $\Weight: E\rightarrow \Reals_{\geq 0}$ and 
nodes $u$, $v$ in the same CSM, what is the weight of the minimum-weight path from $u$ to $v$? 
The problem can be formulated on the tropical semiring $(\Reals_{\geq 0}\cup \{\infty\}, \min, +, \infty,  0)$.
\item \emph{Most probable path} i.e., given a probability function $P: E\rightarrow [0,1]$ and
nodes $u$, $v$ in the same CSM, what is the probability of the highest-probable path from $u$ to $v$? 
The problem can be formulated on the Viterbi semiring $([0,1], \max, \cdot, 0,  1)$.
\item The class of \emph{interprocedural, finite, distributive, subset (IFDS)} problems defined in~\cite{Reps95}.
Given a finite domain $D$, a universe of flow functions $F$ containing distributive functions $f:2^D\rightarrow 2^D$,
a weight function $\Weight:E\rightarrow F$ associates each edge with a flow function. The weight
of an interprocedural path is then defined as the composition $\circ$ of the flow functions along its edges,
and the IFDS problem given nodes $u$, $v$ asks for the meet $\sqcap$ (union or intersection) of the
weights of all $u\rightsquigarrow v$ paths. The problem can be formulated on
the meet-composition semiring $(F, \sqcap, \circ, \emptyset,  I)$, where $I$ is the 
identity function. 
\item The class of \emph{interprocedural distributive environment (IDE)} problems defined in~\cite{Sagiv96}.
This class of dataflow problems is an extension to IFDS, with the difference that the flow functions (called environment transformers)
map elements from the finite domain $D$ to values in an infinite set (e.g., of the form $f:D\rightarrow \Nats$).
An environment transformer is denoted as $f[d\rightarrow \ell]$, meaning that the element $d\in D$ is mapped to value $\ell$,
while the mapping of all other elements remains unchanged. 
The problem can be formulated on
the meet-environment-transformer semiring $(F, \sqcap, \circ, \emptyset,  I)$, where $I$ is the 
identity environment transformer, leaving every map unchanged. 
\end{compactenum}

Note that if we assume that the set of weights of all interprocedural paths
in the system is finite, then the size of this set bounds the stack height $h$.
Additionally, several problems can be formulated as algebraic path problems
in which bounding the stack height 
can be viewed as an approximation to them 
(e.g., shortest path with negative interprocedural cycles, or 
probability of reaching a node $v$ from a node $u$).

\begin{comment}
The following problems can be formulated as algebraic path problems, and bounding the stack height 
can be viewed as an approximation to them.

\begin{compactenum}
\item Shortest path problem with negative weights and possibly negative cycles.
\item The probability of reaching a node $v$ from a node $u$.
\end{compactenum}
\end{comment}

\section{Dynamic Algorithms for Preprocess, Update and Query}\label{sec:dynamic}

In the current section we present algorithms that take as input 
a constant treewidth graph $G$ and a balanced, semi-nice tree-decomposition 
$\Tree(G)$ (recall Theorem~\ref{thm:seminice}), and achieve the following tasks:
\begin{compactenum}
\item Preprocessing the tree-decomposition $\Tree(G)$ of a graph $G$ to answer algebraic path queries fast.
\item Updating the preprocessed $\Tree(G)$ upon change of the weight $\Weight(u,v)$ of an edge $(u,v)$.
\item Querying the preprocessed $\Tree(G)$ to retrieve the distance $\Distance(u,v)$ of any pair of nodes $u,v$.
\end{compactenum}
In the following section we use the results of this section in order to preprocess RSMs fast, with the purpose of answering
interprocedural same-context algebraic path queries fast. Refer to Example~\ref{ex:rsm} of Section~\ref{sec:rsm} for an illustration
on how these algorithms are executed on an RSM.

First we establish the following lemma which captures the main intuition behind tree decompositions,
namely, that bags $\Bag$ of the tree-decomposition $\Tree(G)$ are separators between nodes
of $G$ that belong to disconnected components of $\Tree(G)$ once $\Bag$ is removed.

\smallskip
\begin{lemma}[Separator property]\label{lem:tree_paths}
Consider a graph $G=(V,E)$ and a tree-decomposition $\Tree(G)$.
Let $u,v\in V$, and $P':\Bag_{1}, \Bag_{2},\dots,\Bag_{j}$ be the unique path in $T$ such that $u\in \Bag_{1}$ and $v\in \Bag_{j}$. For each $i\in \{1,\dots, j-1\}$ and for each path $P:u\rightsquigarrow v$, there exists a node $x_{i}\in (\Bag_{i}\cap \Bag_{i+1}\cap P)$.
\end{lemma}
\begin{prf}
Fix a number $i\in \{1,\dots, j-1\}$. We argue that for each path $P:u\rightsquigarrow v$, there exists a node $x_{i}\in (\Bag_{i}\cap \Bag_{i+1}\cap P)$.
We construct a tree $\Tree'(G)$, which is similar to $\Tree(G)$ except that instead of having an edge between bag $\Bag_{i}$ and bag $\Bag_{i+1}$, there is a new bag $\Bag$, that contains the nodes in $\Bag_{i}\cap \Bag_{i+1}$, and there is an edge between $\Bag_{i}$ and $\Bag$ and one between  $\Bag$ and $\Bag_{i+1}$. It is easy to see that $\Tree'(G)$ forms a tree decomposition of $G$. Let $\mathcal{C}_1$, $\mathcal{C}_2$ be the two components of $\Tree(G)$ separated be $\Bag$, and w.l.o.g. $u\in \mathcal{C}_1$ and $v\in\mathcal{C}_2$. It follows by the definition of tree decomposition that $\Bag$ is a separator of $\bigcup_{\Bag'\in \mathcal{C}_1}\Bag'$ and $\bigcup_{\Bag'\in \mathcal{C}_2}\Bag'$.
Hence, each path $u\rightsquigarrow v$ must go through some node $x_i$ in $\Bag$, and by construction $x_i\in \Bag_i\cap \Bag_{i+1}$.
\end{prf}

\smallskip\noindent{\bf Intuition and $\Ushape$-shaped paths.}
A central concept in our algorithms is that of $\Ushape$-shaped paths.
Given a bag $\Bag$ and nodes $u,v\in\Bag$ we say that a path $P:u\rightsquigarrow v$
is $\Ushape$-shaped in $\Bag$, if one of the following conditions hold:
\begin{compactenum}
\item Either $|P|>1$ and for all intermediate nodes $w\in P$, we have $\Level(w)\geq \Level(\Bag)$,
\item or $|P|\leq 1$ and $\Bag$ is $\Bag_u$ or $\Bag_v$. 
\end{compactenum}
Informally, given a bag $\Bag$, a $\Ushape$-shaped path in $\Bag$
is a path that traverses intermediate nodes that are introduced in $\Bag$ and its descendants in $\Tree(G)$.
In the following we present three algorithms for (i)~preprocessing a tree decomposition, (ii)~updating the
data structures of the preprocessing upon a weight change $\Weight(u,v)$ of an edge $(u,v)$, and
(iii)~querying for the distance $\Distance(u,v)$ for any pair of nodes $u,v$. The intuition behind the overall approach
is that for every path $P:u\rightsquigarrow v$ and $z=\argmin_{x\in P}\Level(x)$, the path $P$ can be decomposed
to paths $P_1:u\rightsquigarrow z$ and $P_2:z\rightsquigarrow v$. By Lemma~\ref{lem:tree_paths},
if we consider the path $P':\Bag_u\rightsquigarrow \Bag_z$ and any bag $\Bag_i\in P'$,
we can find nodes $x,y\in \Bag_i\cap P_1$ (not necessarily distinct). Then $P_1$ is decomposed to
a sequence of $\Ushape$-shaped paths $P_1^i$, one for each such $\Bag_i$, and the weight of $P_1$
can be written as the $\otimes$-product of the weights of $P_1^i$, i.e., $\otimes(P_1)=\bigotimes(\otimes(P_1^i))$.
Similar observation holds for $P_2$.
Hence, the task of preprocessing and updating is to summarize in each $\Bag_i$ the weights of all such $\Ushape$-shaped
paths between all pairs of nodes appearing in $\Bag_i$. To answer the query, the algorithm traverses upwards the tree $\Tree(G)$
from $\Bag_u$ and $\Bag_v$, and combines the summarized paths to obtain the weights of all such paths $P_1$ and $P_2$, and eventually
$P$, such that $\otimes(P)=\Distance(u,v)$.

\smallskip\noindent{\bf Informal description of preprocessing.}
Algorithm $\Preprocessalgo$ associates with each bag $\Bag$ a \emph{local distance} map $\LD_{\Bag}:\Bag\times \Bag\rightarrow \Sigma$.
Upon a weight change, algorithm $\Updatealgo$ updates the local distance map of some bags.
It will hold that after the preprocessing and each subsequent update,
$\LD_{\Bag}(u,v) = \bigoplus_{P:u\rightsquigarrow v}\{\otimes(P)\}$, where all $P$ are $\Ushape$-shaped paths in $\Bag$.
Given this guarantee, we later present an algorithm for answering $(u,v)$ queries with $\Distance(u,v)$, the 
distance from $u$ to $v$. Algorithm $\Preprocessalgo$ is a dynamic programming algorithm. 
It traverses $\Tree(G)$ bottom-up, and for a currently examined bag $\Bag$ introducing a node $x$, it calls the method $\Mergealgo$ to compute 
the local distance map $\LD_{\Bag}$. In turn, $\Mergealgo$ computes $\LD_{\Bag}$ depending only on the 
local distance maps $\LD_{\Bag_i}$ of the children $\{\Bag_i\}$ of $\Bag$, and uses the closure operator $*$ to capture
possibly unbdounded traversals of cycles whose smallest-level node is $x$.
 See Method~\ref{algo:merge} and Algorithm~\ref{algo:preprocess} for a formal description.

\smallskip
\begin{algorithm}%[H]
\small
\SetAlgoNoLine
\SetAlgorithmName{Method}{method}
\DontPrintSemicolon
%\setstretch{1.05}
\caption{$\Mergealgo$}\label{algo:merge}
\KwIn{A bag $\Bag_x$ with children $\{\Bag_i\}$}
\KwOut{A local distance map $\LD_{\Bag_x}$}
\BlankLine
\label{line:all_cycles}Assign $\Weight'(x,x)\leftarrow \left(\bigotimes\{\LD_{\Bag_1}(x,x)^{\ast},\dots, \LD_{\Bag_j}(x,x)^{\ast}\}\right)^{\ast}$\\
\ForEach{$u\in \Bag_x$ with $u\neq x$}{
\label{line:xu}Assign $\Weight'(x,u) \leftarrow \bigoplus\{\Weight(x,u), \LD_{\Bag_1}(x,u),\dots, \LD_{\Bag_j}(x,u)\}$\\
\label{line:ux}Assign $\Weight'(u,x) \leftarrow \bigoplus\{\Weight(u,x), \LD_{\Bag_1}(u,x),\dots, \LD_{\Bag_j}(u,x)\}$\\
}
\ForEach{$u, v\in \Bag_x$}{
\label{line:delta}Assign $\delta\leftarrow\bigotimes(\Weight'(u,x), \Weight'(x,x), \Weight'(x,v))$\\
%Assign $\delta\leftarrow (\Weight(u,x)^{\ast}\otimes\Weight(x,v)^{\ast})^{\ast}$\\
Assign $\LD_{\Bag_x}(u,v)\leftarrow \bigoplus\{\delta, \LD_{\Bag_1}(u,v),\dots, \LD_{\Bag_j}(u,v)\}$\\
\label{line:merge_children}
}
\end{algorithm}
 
\smallskip
\begin{algorithm}%[H]
\small
\SetAlgoNoLine
\DontPrintSemicolon
%\setstretch{1.05}
\caption{$\Preprocessalgo$}\label{algo:preprocess}
\KwIn{A tree-decomposition $\Tree(G)=(V_T, E_T)$}
\KwOut{A local distance map $\LD_{\Bag}$ for each bag $\Bag\in V_T$}
\BlankLine
Traverse $\Tree(G)$ bottom up and examine each bag $\Bag$ with children $\{\Bag_i\}$\\
\eIf{$\Bag$ introduces some node $x$}{
Assign $\LD_{\Bag}\leftarrow \Mergealgo$ on $\Bag$
}
{
\ForEach{$u,v\in \Bag$}{
Assign $\LD_{\Bag}(u,v)\leftarrow \bigoplus\{\LD_{\Bag_1}(u,v),\dots, \LD_{\Bag_j}(u,v)\}$\\
}
}
\end{algorithm}

\begin{lemma}\label{lem:preprocess_correctness}
%Consider a graph $G=(V,E)$ and a tree-decomposition $\Tree(G)$.
At the end of $\Preprocessalgo$, for every bag $\Bag$ and nodes $u,v\in \Bag$,
we have $\LD_{\Bag}(u,v)=\bigoplus_{P:u\rightsquigarrow v}\{\otimes(P)\}$,
where all $P$ are $\Ushape$-shaped paths in $\Bag$.
\end{lemma}
\begin{prf}
The proof is by induction on the parents.
Initially, $\Bag$ is a leaf introducing some node $x$, thus each such path $P$ can only go through $x$, and hence will
be captured by $\Preprocessalgo$.
Now assume that the algorithm examines a bag $\Bag$, and by the induction hypothesis the statement is true for all $\{\Bag_i\}$ children of $\Bag_x$.
The correctness follows easily if $\Bag$ does not introduce any node, since every such $P$ is a $\Ushape$-shaped path in
some child $\Bag_i$ of $\Bag$. 
Now consider that $\Bag$ introduces some node $x$, and any $\Ushape$-shaped path $P':u\rightsquigarrow v$ that additionally visits $x$,
and decompose it to paths $P_1:u\rightsquigarrow x$, $P_2:x\rightsquigarrow x$ and $P_3:x\rightsquigarrow v$,
such that $x$ is not an intermediate node in either $P_1$ or $P_3$, and we have
by distributivity:
\begin{align*}
\bigoplus_{P'}\otimes(P')&=\bigoplus_{P_1, P_2, P_3}\bigotimes\left(\otimes(P_1),\otimes(P_2),\otimes(P_3)\right)\\
&=\bigotimes\left(\bigoplus_{P_1}\otimes(P_1), \bigoplus_{P_2}\otimes(P_2), \bigoplus_{P_3}\otimes(P_3)\right)
\end{align*}
Note that $P_1$ and $P_3$ are also $\Ushape$-shaped in one of the children
bags $\Bag_i$ of $\Bag_x$, hence by the induction hypothesis in lines \ref{line:ux} and \ref{line:xu} of $\Mergealgo$ we have
$\Weight'(u,x)=\bigoplus_{P_1}\otimes(P_1)$ and $\Weight'(x,v)=\bigoplus_{P_3}\otimes(P_3)$.
Also, by decomposing $P_2$ into a (possibly unbounded) sequence of paths $P_2^i:x\rightsquigarrow x$ such
that $x$ is not intermediate node in any $P_2^i$, we get that each such $P_2^i$ is a $\Ushape$-shaped
path in some child $\Bag_{l_i}$ of $\Bag$, and we have by distributivity and the induction hypothesis
\begin{align*}
\bigoplus_{P_2}\otimes(P_2) &= \bigoplus_{P_2^1,P_2^2,\dots}\bigotimes\left\{\otimes(P_2^1), \otimes(P_2^2),\dots\right\}\\
&=\bigoplus_{\Bag_{l_1},\Bag_{l_2},\dots} \bigotimes\left\{\bigoplus_{P_2^1}\otimes(P_2^1), \bigoplus_{P_2^2}\otimes(P_2^2),\dots\right\}\\
&=\bigoplus_{\Bag_{l_1},\Bag_{l_2},\dots} \bigotimes\left\{\LD_{\Bag_{l_1}}(x,x), \LD_{\Bag_{l_2}}(x,x),\dots\right\}
\end{align*}

and the last expression equals $\Weight'(x,x)$ from line \ref{line:all_cycles} of $\Mergealgo$.
The above conclude that in line \ref{line:delta} of $\Mergealgo$ we have $\delta=\bigoplus_{P'}\otimes(P')$.

\begin{comment}
It follows from the definitions that each $P_i$ is either a $\Ushape$-shaped path
in one of the children $\Bag_i$, or $|P_i|\leq 1$.
By associativity of $\otimes$, it is $\otimes(P')=\bigotimes(\otimes(P_1), \otimes(P_2))$,
and hence we have 
\begin{align*}
\bigoplus_{P'}\otimes(P')&=\bigoplus_{P_1, P_2}\bigotimes\left(\otimes(P_1),\otimes(P_2)\right)\\
&=\bigotimes\left(\bigoplus_{P_1}\otimes(P_1), \bigoplus_{P_2}\otimes(P_2)\right)
\end{align*}
by factoring together all paths $P_i$, because of distributivity.
By the induction hypothesis we have that
$\Weight'(u,x)=\bigoplus_{P_1}\otimes(P_1)$ and $\Weight'(x,v)=\bigoplus_{P_2}\otimes(P_2)$.
It follows that $\delta=\bigoplus_{P':u\rightsquigarrow v}\{\otimes(P')\}$.
\end{comment}
Finally, each $\Ushape$-shaped path $P:u\rightsquigarrow v$ in $\Bag$ either visits $x$, or is $\Ushape$-shaped 
in one of the children $\Bag_i$. Hence after line~\ref{line:merge_children} of Method $\Mergealgo$
has run on $\Bag$,  for all $u,v\in \Bag$ we have that $\LD_{\Bag}(u,v)=\bigoplus_{P:u\rightsquigarrow v}\otimes(P)$
where all paths $P$ are $\Ushape$-shaped in $\Bag$. The desired results follows.
\end{prf}

\begin{lemma}\label{lem:preprocess_complexity}
%Consider a graph $G=(V,E)$ of $n$ nodes and a tree-decomposition $\Tree(G)$
%%Algorithm 
$\Preprocessalgo$ requires $O(n)$ semiring operations.
\end{lemma}
\begin{prf}
%%Method 
$\Mergealgo$ requires $O(t^2)=O(1)$ operations, and $\Preprocessalgo$ calls $\Mergealgo$ at most
once for each bag, hence requiring $O(n)$ operations.
\end{prf}

\begin{figure}[!h]
\centering
\begin{tikzpicture}[thick, >=latex,
pre/.style={<-,shorten >= 1pt, shorten <=1pt, thick},
post/.style={->,shorten >= 1pt, shorten <=1pt,  thick},
und/.style={very thick, draw=gray},
node/.style={circle, minimum size=2mm, draw=black!100, line width=1pt, inner sep=3},
rootbag/.style={ellipse, minimum height=7mm,minimum width=14mm,draw=black!80, line width=2.5pt},
virt/.style={circle,draw=black!50,fill=black!20, opacity=0}]

\node	[node]		(u)	at	(-3, 0)	{$u$};
\node	[virt]		(u2)	at	(-3, -0.3)	{};
\node	[node]		(x)	at	(0, -1)	{$x$};
\node	[virt]		(x3)	at	(-0.2, -1.6)	{};
\node	[virt]		(x4)	at	(0.2, -1.3)	{};
\node	[node]		(v)	at	(3, 0.6)	{$v$};
\node	[virt]		(v2)	at	(3, -0.05)	{};

\node	[virt]		(y3)	at	(-0.05, -1.3)	{};
\node	[virt]		(y4)	at	(0.05, -1.6)	{};

\node	[]		(P1)	at	(-1.5, -2.1)	{$P_1$};
\node	[]		(P2)	at	(1.5, -1.8)	{$P_3$};
\node	[]		(P2)	at	(0, -2.3)	{$P_2$};

\node	[]		(a1)	at	(-2.6, -1.9)	{};
\node	[]		(a2)	at	(-0.4,-2.1)	{};
\node	[]		(a3)	at	(-1.5,-2.5)	{};

\node	[]		(b2)	at	(2.6, -1.2)	{};
\node	[]		(b1)	at	(0.4,-1.9)	{};
\node	[]		(b3)	at	(1.5,-2.2)	{};

\node	[]		(c2)	at	(0.35, -2.4)	{};
\node	[]		(c1)	at	(-0.35,-2.4)	{};
\node	[]		(c3)	at	(0,-2.6)	{};

\draw [draw=black, line width =1, ->, dashed, dash pattern=on 2mm off 1mm] plot [smooth, ] coordinates {(u2) (a1) (a3) (a2) (x3) };
\draw [draw=black, line width =1,dashed, dash pattern=on 2mm off 1mm, ->] plot [smooth, ] coordinates {(x4) (b1) (b3) (b2) (v2) };
\draw [draw=black, line width =1,dashed, dash pattern=on 2mm off 1mm, ->] plot [smooth, ] coordinates {(y3) (c1) (c3)  (c2) (y4) };

\draw [draw=black, line width =1, ->] (u) to node[above=0.2] {$\Weight'(u,x)$} (x);
\draw [draw=black, line width =1, ->] (x) to node[above=0.2]{$\Weight'(x,v)$} (v);
\draw [draw=black, loop above, looseness=20, line width =1, ->] (x) to node[above=0.2]{$\Weight'(x,x)$} (x);

\end{tikzpicture}
\caption{Illustration of the inductive argument of $\Preprocessalgo$.
%When examining bag $\Bag_x$, any $\Ushape$-shaped path $P':u\rightsquigarrow v$
%that goes through $x$ can be decomposed into paths $P_1:u\rightsquigarrow x$ and
%$P_2:x\rightsquigarrow v$, that are $\Ushape$-shaped in some $\Bag_i$ children of $\Bag_x$,
%and by the induction hypothesis $\Weight'(u,x)=\otimes(P_1)$ and $\Weight'(x,v)=\otimes(P_2)$.
%Hence, when $\Preprocessalgo$ finishes processing $\Bag_x$ it will be $\LD_{\Bag_x}(u,v)=\bigoplus_{P}\otimes(P)$,
%where $P$ are $\Ushape$-shaped paths in $\Bag_x$.
}\label{fig:preprocess}
\end{figure}
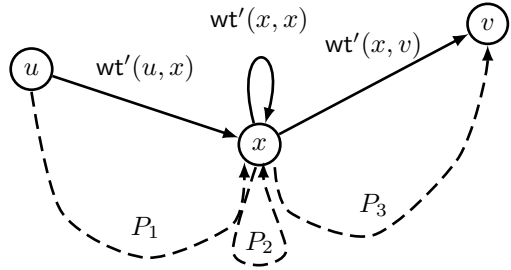

\smallskip\noindent{\bf Informal description of updating.}
Algorithm $\Updatealgo$ is called whenever the weight $\Weight(x,y)$ of an edge of $G$
has changed. Given the guarantee of Lemma~\ref{lem:preprocess_correctness}, after $\Updatealgo$
has run on an edge update $\Weight(x,y)$, it restores the property that for each bag $\Bag$
we have $\LD_{\Bag}(u,v)=\bigoplus_{P:u\rightsquigarrow v}\{\otimes(P)\}$,
where all $P$ are $\Ushape$-shaped paths in $\Bag$. See Algorithm~\ref{algo:update} for a formal description.

\smallskip
\begin{algorithm}%[H]
\small
\SetAlgoNoLine
\DontPrintSemicolon
%\setstretch{1.05}
\caption{$\Updatealgo$}\label{algo:update}
\KwIn{An edge $(x,y)$ with new weight $\Weight(x,y)$}
\KwOut{A local distance map $\LD_{\Bag}$ for each bag $\Bag\in V_T$}
\BlankLine
Assign $\Bag\leftarrow \Bag_{(x,y)}$, the highest bag containing the edge $(x,y)$\\
\Repeat{$\Level(\Bag)=0$}{
Call $\Mergealgo$ on $\Bag$\\
Assign $\Bag\leftarrow \Bag'$ where $\Bag'$ is the parent of $\Bag$\\
}
\end{algorithm}

\begin{lemma}\label{lem:update_correctness}
%Consider a graph $G=(V,E)$ and a tree-decomposition $\Tree(G)$.
At the end of each run of $\Updatealgo$, for every bag $\Bag$ and nodes $u,v\in \Bag$,
we have $\LD_{\Bag}(u,v)=\bigoplus_{P:u\rightsquigarrow v}\{\otimes(P)\}$,
where all $P$ are $\Ushape$-shaped paths in $\Bag$.
\end{lemma}
\begin{prf}
First, by the definition of a $\Ushape$-shaped path $P$ in $\Bag$ it follows
that the statement holds for all bags not processed by $\Updatealgo$,
since for any such bag $\Bag$ and $\Ushape$-shaped path $P$ in $\Bag$,
the path $P$ cannot traverse $(u,v)$.
For the remaining bags,
the proof follows an induction on the parents
updated by $\Updatealgo$, similar to that of Lemma~\ref{lem:preprocess_correctness}.
\end{prf}

\begin{lemma}\label{lem:update_complexity}
%Consider a graph $G=(V,E)$ and a tree-decomposition $\Tree(G)$.
%%Algorithm 
$\Updatealgo$ requires $O(\log n)$ operations per update.
\end{lemma}
\begin{prf}
$\Mergealgo$ requires $O(t^2)=O(1)$ operations, and $\Updatealgo$ calls
$\Mergealgo$ once for each bag in the path from $\Bag_{(u,v)}$ to the root. 
Recall that the height of $\Tree(G)$ is $O(\log n)$ (Theorem~\ref{thm:seminice}), and the result follows.
\end{prf}

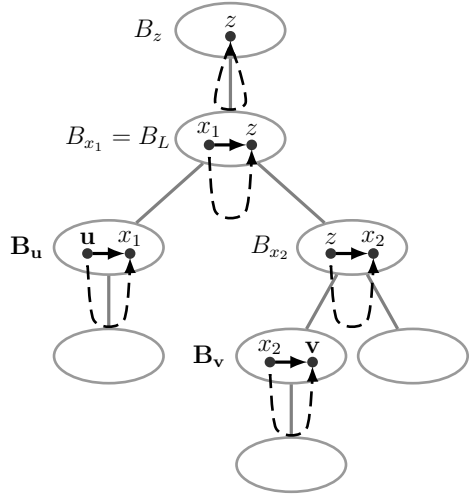
\begin{figure}[!h]
\centering
\large
\begin{tikzpicture}[thick, >=latex, scale=0.8,every node/.style={transform shape},
pre/.style={<-,shorten >= 1pt, shorten <=1pt, thick},
post/.style={->,shorten >= 1pt, shorten <=1pt,  thick},
und/.style={very thick, draw=gray},
bag/.style={ellipse, minimum height=9mm,minimum width=18mm,draw=gray!80, line width=1pt},
rootbag/.style={ellipse, minimum height=7mm,minimum width=14mm,draw=black!80, line width=2.5pt},
node/.style={circle,draw=black!80, fill=black!80, inner sep=0, minimum size=4pt},
virt/.style={circle,draw=black!50,fill=black!20, opacity=0}]

\newcommand{\xdistone}{-0.35}
\newcommand{\xdisttwo}{0.35}

\node	[bag]		(B0)	at	(0, 0)				{};
\node	[bag]		(B1)	at	(0, -1.8)		{};
\node	[bag]		(B2)	at	(-2, -3.6)		{};
\node	[bag]		(B3)	at	(-2, -5.4)	{};
\node	[bag]		(B4)	at	(2, -3.6)			{};
\node	[bag]		(B5)	at	(1, -5.4)		{};
\node	[bag]		(B6)	at	(3, -5.4)		{};
\node	[bag]		(B7)	at	(1, -7.2)			{};

\draw [und] (B0) to (B1);
\draw [und] (B1) to (B2);
\draw [und] (B2) to (B3);
\draw [und] (B1) to (B4);
\draw [und] (B4) to (B5);
\draw [und] (B5) to (B7);
\draw [und] (B4) to (B6);

\node	[node]		(x1)		at	(-2+\xdistone, -3.7)		{};
\node	[node]		(x2)		at	(-2+\xdisttwo, -3.7)		{};
\node	[]			(us)		at	(-2+\xdistone, -3.45)	{$\mathbf{u}$};
\node	[]			(Bu)		at	(-2+\xdistone - 1, -3.6)	{$\mathbf{\Bag_u}$};
\node	[]			(x2s)		at	(-2+\xdisttwo, -3.45)	{$x_1$};

\node	[node]		(x3)		at	(0+\xdistone, -1.9)		{};
\node	[node]		(x4)		at	(0+\xdisttwo, -1.9)		{};
\node	[]			(x3s)		at	(0+\xdistone, -1.65)	{$x_1$};
\node	[]			(x4s)		at	(0+\xdisttwo, -1.65)	{$z$};
\node	[]			(Bx1)		at	(-0.5+\xdistone - 1, -1.8)	{$\Bag_{x_1}=\Bag_L$};

\node	[node]		(x5)		at	(2+\xdistone, -3.7)		{};
\node	[node]		(x6)		at	(2+\xdisttwo, -3.7)		{};
\node	[]			(x5s)		at	(2+\xdistone, -3.45)	{$z$};
\node	[]			(x6s)		at	(2+\xdisttwo, -3.45)	{$x_2$};
\node	[]			(Bu)		at	(2+\xdistone - 1, -3.6)	{$\Bag_{x_2}$};

\node	[node]		(x7)		at	(1+\xdistone, -5.5)		{};
\node	[node]		(x8)		at	(1+\xdisttwo, -5.5)		{};
\node	[]			(vs)		at	(1+\xdisttwo, -5.25)		{$\mathbf{v}$};
\node	[]		(x7s)		at	(1+\xdistone, -5.25)		{$x_2$};
\node	[]		(x8s)		at	(1+\xdistone - 1, -5.4)	{$\mathbf{\Bag_v}$};
\node	[]			(Bz)		at	(0+\xdistone - 1,0)	{$\Bag_z$};

\node	[node]		(z)		at	(0, -0.1)		{};
\node	[]		(zs)		at	(0, 0.15)		{$z$};
%\node	[]		(x8s)		at	(-1.05+\xdisttwo+0.7, -5.4)		{$\Bag_v$};

\node	[node, opacity=0]		(y1)		at	(-1.9+\xdistone, -4.8)		{};
\node	[node, opacity=0]		(y2)		at	(-2.1+\xdisttwo, -4.8)		{};

\node	[node, opacity=0]		(y3)		at	(0.1+\xdistone, -3)		{};
\node	[node, opacity=0]		(y4)		at	(-0.1+\xdisttwo, -3)		{};

\node	[node, opacity=0]		(y5)		at	(2.1+\xdistone, -4.8)		{};
\node	[node, opacity=0]		(y6)		at	(1.9+\xdisttwo, -4.8)		{};

\node	[node, opacity=0]		(y7)		at	(1.1+\xdistone, -6.6)		{};
\node	[node, opacity=0]		(y8)		at	(0.9+\xdisttwo, -6.6)		{};

\node	[node, opacity=0]		(yz1)		at	(0.1+\xdistone, -1.2)		{};
\node	[node, opacity=0]		(yz2)		at	(-0.1+\xdisttwo, -1.2)		{};

\draw [->, very thick] (x1) to  node [auto, above=0.4] {} 	(x2);
\draw [->, very thick] (x3) to  node [auto, below=0.4] {} 	(x4);
\draw [->, very thick] (x5) to  node [auto, above=0.5] {} 	(x6);
\draw [->, very thick] (x7) to  node [auto, above=0.5] {} 	(x8);

\draw [draw=black, dashed, line width =1, ->, dash pattern=on 2mm off 1mm, shorten >= -12pt, shorten <=-5pt,] plot [smooth, ] coordinates {(x1) (y1) (y2) (x2) };
\draw [draw=black, dashed, line width =1, ->, dash pattern=on 2mm off 1mm, shorten >= -12pt, shorten <=-5pt,] plot [smooth, ] coordinates {(x3) (y3) (y4) (x4) };
\draw [draw=black, dashed, line width =1, ->, dash pattern=on 2mm off 1mm, shorten >= -12pt, shorten <=-5pt,] plot [smooth, ] coordinates {(x5) (y5) (y6) (x6) };
\draw [draw=black, dashed, line width =1, ->, dash pattern=on 2mm off 1mm, shorten >= -12pt, shorten <=-5pt,] plot [smooth, ] coordinates {(x7) (y7) (y8) (x8) };
\draw [draw=black, dashed, line width =1, ->, dash pattern=on 2mm off 1mm, shorten >= -12pt, shorten <=-5pt,] plot [smooth, ] coordinates {(z) (yz1) (yz2) (z) };

%\draw [draw=black, dashed, line width =1, ->, dash pattern=on 2mm off 1mm, shorten >= -12pt, shorten <=-5pt,] plot [smooth, ] coordinates {(x1) (y1) (z1) (z2) (y2) (x2) (x3) (y3) (z3) (z4) (y4) (x4) (x6) (x5) (x7) (y5) (z5) (z6) (y6) (x8)  };
\end{tikzpicture}
\caption{Illustration of $\Queryalgo$ in computing the distance $\Distance(u,v)=\otimes(P)$ as a sequence of $\Ushape$-shaped paths,
whose weight has been captured in the local distance map of each bag.
When $\Bag_z$ is examined, with $z=\argmin_{x\in P}\Level(x)$, it will be $\delta_u(z)=\Distance(u,z)$ and $\delta_v(z)=\Distance(z,v)$, and hence by distributivity
$\Distance(u,v)=\bigotimes(\delta_u(z), \delta_v(z))$.}\label{fig:query}
\end{figure}

\noindent{\bf Informal description of querying.}
Algorithm $\Queryalgo$ answers a $(u,v)$ query with the distance $\Distance(u,v)$ from $u$ to $v$.
Because of Lemma~\ref{lem:tree_paths}, every path $P:u\rightsquigarrow v$ is guaranteed to go
through the least common ancestor (LCA) $\Bag_L$ of $\Bag_u$ and $\Bag_v$, and possibly some of the ancestors $\Bag$ of $\Bag_L$.
Given this fact, algorithm $\Queryalgo$ uses the procedure $\Climbalgo$ to climb up the tree from $\Bag_u$ and $\Bag_v$
until it reaches $\Bag_L$ and then the root of $\Tree(G)$. For each encountered bag $\Bag$ along the way,
it computes maps $\delta_u(w)=\bigoplus_{P_1}\{\otimes(P_1)\}$, and $\delta_v(w)=\bigoplus_{P_2}\{\otimes(P_2)\}$ 
where all $P_1:u\rightsquigarrow w$ and $P_2:w\rightsquigarrow v$ are such that each intermediate node $y$ in them has been introduced in $\Bag$.
This guarantees that for path $P$ such that $\Distance(u,v)=\otimes(P)$, when $\Queryalgo$ examines
the bag $\Bag_z$ introducing $z=\argmin_{x\in P}\Level(x)$, it will be $\Distance(u,v)=\bigotimes(\delta_u(z),\delta_v(z))$.
Hence, for $\Queryalgo$ it suffices to maintain a current best solution $\delta$, and update it with $\delta\leftarrow  \bigoplus\{\delta,\bigotimes(\delta_u(x), \delta_v(x))\}$
every time it examines a bag $\Bag$ introducing some node $x$. Figure~\ref{fig:query}
presents a pictorial illustration of $\Queryalgo$ and its correctness.
Method~\ref{algo:climb} presents the $\Climbalgo$ procedure which,
given a current distance map of a node $\delta$, a current bag $\Bag$ and a flag $\Up$, updates $\delta$
with the distance to (if $\Up=\True$), or from (if $\Up=\False$) each node in $\Bag$.
See Method~\ref{algo:climb} and Algorithm~\ref{algo:query} for a formal description.

\smallskip
\begin{algorithm}%[H]
\small
\SetAlgoNoLine
\SetAlgorithmName{Method}{method}
\DontPrintSemicolon
%\setstretch{1.05}
\caption{$\Climbalgo$}\label{algo:climb}
\KwIn{A bag $\Bag$, a map $\delta$, a flag $\Up$}
\KwOut{A new map $\delta$}
\BlankLine
Remove from $\delta$ all $w\not\in \Bag$\\
Assign $\delta(w)\leftarrow\Zero$ for all $w\in \Bag$ and not in $\delta$\\
\uIf{$\Bag$ introduces node $x$}{
\eIf(\tcc*[f]{Climbing up}){$\Up$}{
Update $\delta$ with $\delta(w)\leftarrow\bigoplus\{\delta(w),\bigotimes(\delta(x), \LD_{\Bag}(x,w))\}$\\
}(\tcc*[f]{Climbing down})
{
Update $\delta$ with $\delta(w)\leftarrow\bigoplus\{\delta(w),\bigotimes(\delta(x), \LD_{\Bag}(w,x))\}$\\
}
}
\Return{$\delta$}
\end{algorithm}

\smallskip
\begin{algorithm}%[H]
\small
\SetAlgoNoLine
\DontPrintSemicolon
%\setstretch{1.05}
\caption{$\Queryalgo$}\label{algo:query}
\KwIn{A pair $(u,v)$}
\KwOut{The distance $\Distance(u,v)$ from $u$ to $v$}
\BlankLine
Initialize map $\delta_u$ with $\delta_u(w)\leftarrow\LD_{\Bag_u}(u,w)$\label{line:init_u}\\
Initialize map $\delta_v$ with $\delta_v(w)\leftarrow\LD_{\Bag_v}(w,v)$\\
Assign $\Bag_L\leftarrow$ the LCA of $\Bag_u$, $\Bag_v$ in $\Tree(G)$\\
Assign $\Bag\leftarrow \Bag_u$\\
\Repeat{$\Bag=\Bag_L$}{
Assign $\Bag\leftarrow \Bag'$ where $\Bag'$ is the parent of $\Bag$\\
Call $\Climbalgo$ on $\Bag$ and $\delta_u$ with flag $\Up$ set to $\True$\label{line:climb1}\\
}
Assign $\Bag\leftarrow \Bag_v$\\
\Repeat{$\Bag=\Bag_L$}{
Assign $\Bag\leftarrow \Bag'$ where $\Bag'$ is the parent of $\Bag$\\
Call $\Climbalgo$ on $\Bag$ and $\delta_v$ with flag $\Up$ set to $\False$\\
}
Assign $\Bag\leftarrow \Bag_L$\\
Assign $\delta\leftarrow  \bigoplus_{x\in \Bag_L} \otimes(\delta_u(x), \delta_v(x))$\\
\Repeat{$\Level(\Bag)=0$}{
Assign $\Bag\leftarrow \Bag'$ where $\Bag'$ is the parent of $\Bag$\\
Call $\Climbalgo$ on $\Bag$ and $\delta_u$ with flag $\Up$ set to $\True$\label{line:climb2}\\
Call $\Climbalgo$ on $\Bag$ and $\delta_v$ with flag $\Up$ set to $\False$\\
\uIf{$\Bag$ introduces node $x$}{
Assign $\delta\leftarrow  \bigoplus\{\delta,\bigotimes(\delta_u(x), \delta_v(x))\}$\\
}
}
\Return{$\delta$}
\end{algorithm}

\smallskip
\begin{lemma}\label{lem:query_correctness}
%Consider a graph $G=(V,E)$ and a tree-decomposition $\Tree(G)$.
$\Queryalgo$ returns $\delta=\Distance(u,v)$.
\end{lemma}
\begin{prf}
Let $P:u\rightsquigarrow v$ be any path from $u$ to $v$, and $z=\argmin_{x\in P}\Level(x)$ the
lowest level node in $P$. Decompose $P$ to $P_1:u\rightsquigarrow z$, $P_2:z\rightsquigarrow v$, and it follows
that $\otimes(P)=\bigotimes(\otimes(P_1),\otimes(P_2))$. We argue that when $\Queryalgo$ examines
$\Bag_z$, it will be $\delta_u(z)=\bigoplus_{P_1}\otimes(P_1)$ and $\bigoplus_{P_2}\delta_v(z)=\otimes(P_2)$. We only focus on the $\delta_u(z)$
case here, as the $\delta_v(z)$ is similar. We argue inductively that when algorithm $\Queryalgo$ examines a bag $\Bag_x$,
for all $w\in \Bag_x$ we have $\delta_u(w)=\bigoplus_{P'}\{\otimes(P')\}$, where all $P'$ are such that for each intermediate node
$y$ we have $\Level(y)\geq \Level(x)$. Initially (line~\ref{line:init_u}), it is $x=u$, $\Bag_x=\Bag_u$, and every such $P'$ is $\Ushape$-shaped in $\Bag_u$,
hence $\LD_{\Bag_x}(x,w)=\bigoplus_{P'}\{\otimes(P')\}$ and $\delta_u(w)=\bigoplus_{P'}\{\otimes(P')\}$.
Now consider that $\Queryalgo$ examines a bag $\Bag_x$ (Lines~\ref{line:climb1} and~\ref{line:climb2})
and the claim holds for $\Bag_{x'}$ a descendant of $\Bag_x$ previously examined by $\Queryalgo$. 
If $x$ does not occur in $P'$, it is a consequence of Lemma~\ref{lem:tree_paths}
that $w\in \Bag_{x'}$, hence by the induction hypothesis, $P'$ has been considered by $\Queryalgo$. Otherwise,
$x$ occurs in $P'$ and decompose $P'$ to $P'_1$, $P'_2$, such that $P'_1$ ends with the first occurrence of $x$ in $P'$,
and it is $\otimes(P)=\bigotimes(\otimes(P'_1),\otimes(P'_2))$. Note that $P'_2$ is a $\Ushape$-shaped path in $\Bag_x$,
hence $\LD_{\Bag_x}(x,w)=\bigoplus_{P'_2}\{\otimes(P'_2)\}$. Finally, as a consequence of Lemma~\ref{lem:tree_paths},
we have that $x\in \Bag_{x'}$, and by the induction hypothesis, $\delta_u(x)=\bigoplus_{P'_1}\{\otimes(P'_1)\}$.
It follows that after $\Queryalgo$ processes $\Bag_x$, it will be $\delta_u(w)=\bigoplus_{P'}\{\otimes(P')\}$.
By the choice of $z$, when $\Queryalgo$ examines the bag $\Bag_z$, it will be $\delta_u(z)=\bigoplus_{P_1}\{\otimes(P_1)\}$.
A similar argument shows that at that point it will also be $\delta_v(z)=\bigoplus_{P_2}\{\otimes(P_2)\}$, hence 
at that point $\delta=\bigotimes(\otimes(P_1),\otimes(P_2))=\Distance(u,v)$.
\end{prf}

\begin{lemma}\label{lem:query_complexity}
%Consider a graph $G=(V,E)$ and a tree-decomposition $\Tree(G)$.Algorithm 
$\Queryalgo$ requires $O(\log n)$ semiring operations.
\end{lemma}
\begin{prf}
%%It is straightforward to see that 
$\Climbalgo$ requires $O(t^2)=O(1)$ operations and $\Queryalgo$ calls $\Climbalgo$ once for 
every bag in the paths from $\Bag_u$ and $\Bag_v$ to the root.
Recall that the height of $\Tree(G)$ is $O(\log n)$ (Theorem~\ref{thm:seminice}), and the result follows.
\end{prf}

We conclude the results of this section with the following theorem.
\smallskip
\begin{theorem}\label{thm:dynamization}
Consider a graph $G=(V,E)$ and a balanced, semi-nice tree-decomposition $\Tree(G)$ of constant treewidth. 
The following assertions hold:
\begin{compactenum}
\item $\Preprocessalgo$ requires $O(n)$ semiring operations;
\item $\Updatealgo$ requires $O(\log n)$ semiring operations per edge weight update; and
\item $\Queryalgo$ correctly answers distance queries in $O(\log n)$ semiring operations.
\end{compactenum}
\end{theorem}

\smallskip\noindent{\bf Witness paths.}
Our algorithms so far have only been concerned with returning the distance $\Distance(u,v)$ of the pair query $u,v$.
When the semiring lacks the closure operator (i.e., for all $s\in \Sigma$ it is $s^*=\One$), as in most problems
e.g., reachability and shortest paths with positive weights, the distance from every $u$ to $v$ is realized by an acyclic path.
Then, it is straightforward to also obtain a witness path, i.e., a path $P:u\rightsquigarrow v$ such that $\otimes(P)=\Distance(u,v)$,
with some minor additional preprocessing. 
Here we outline how.

Whenever $\Mergealgo$ updates the local distance $\LD_{\Bag}(u,v)$ between
two nodes in a bag $\Bag$, it does so by considering the distances to and from an intermediate node $x$.
It suffices to remember that intermediate node for every such local distance. Then, the witness path to a
local distance in $\Bag$ can be obtained straightforwardly by a top-down computation on $\Tree(G)$
starting from $\Bag$.
Recall that in essence, $\Queryalgo$ answers a distance query $u,v$ by combining several local distances
along the paths $\Bag_u\rightsquigarrow \Bag_z$ and $\Bag_z\rightsquigarrow \Bag_v$, where $z$ is the node
with the minimum level in a path $P:u\rightsquigarrow v$ such that $\otimes(P)=\Distance(u,v)$. Since from
every such local distance a witness sub-path $P_i$ can be obtained, $P$ is reconstructed
by juxtaposition of all such $P_i$. Finally, this process costs $O(|P|)$ time.

\section{Algorithms for Constant Treewidth RSMs}\label{sec:rsm}

In this section we consider the bounded height algebraic path problem on RSMs of constant treewidth. That is, we consider (i)~an RSM $A=\set{A_1,A_2,\dots, A_k}$, where $A_i$ consists of $n_i$ nodes and $b_i$ boxes; 
(ii)~a partially complete semiring $(\Sigma, \oplus, \otimes, \Zero, \One)$; and (iii)~a maximum stack height $h$. Our task is to create a datastructure that after some preprocessing can answer queries of the form: Given a pair $((u,\emptyset),(v,\emptyset))$ of configurations  compute $\Distance((u,\emptyset),(v,\emptyset),h)$ (also recall Remark~\ref{rem:same_context}).
For this purpose, we present the algorithm $\RMSDist$,
which performs such preprocessing using a datastructure $\Datastructure$ consisting of the algorithms $\Preprocessalgo$, $\Updatealgo$ and $\Queryalgo$ of Section~\ref{sec:dynamic}. At the end of $\RMSDist$ it will hold that algebraic path pair queries in a CSM $A_i$ can be answered in $O(\log n_i)$ semiring operations.
We later present some additional preprocessing which suffers a factor of $O(\log n_i)$ in the preprocessing space, but reduces the pair query time to constant.

\smallskip\noindent{\bf Algorithm $\RMSDist$.}
Our algorithm $\RMSDist$ can be viewed as %an instance of the value iteration paradigm from algorithmic game theory. 
a Bellman-Ford computation on the call graph of the RSM (i.e., a graph where every node corresponds to a CSM, and an edge connects two CSMs if one appears as a box in the other). 
Informally, $\RMSDist$ consists of the following steps.
\begin{compactenum}
\item First, it preprocesses the control flow graphs $G_i=(V_i,E_i')$ of the CSMs $A_i$ using $\Preprocessalgo$ of Section~\ref{sec:dynamic}, where the weight function $\Weight_i$ for each $G_i$ is extended such that $\Weight_i((en,b),(ex,b))=\Zero$ for all pairs of call and return nodes to the same box $b$. This allows the computation of $\Distance(u,v,0)$ for all pairs of nodes $(u,v)$, since no call can be made while still having zero stack height.
\item Then, iteratively for each $\ell$, where $0\leq \ell\leq h-1$, given that we have a dynamic datastructure $\Datastructure$ (concretely, an instance of the dynamic algorithms $\Updatealgo$ and $\Queryalgo$ from Section~\ref{sec:dynamic}) for computing $\Distance(u,v,\ell)$, the algorithm does as follows: First, for each $G_i$ whose entry to exit distance $\Distance(En_i,Ex_i)$ has changed from the last iteration and for each $G_j$ that contains a box pointing to $G_i$, it updates the call to return distance of the corresponding nodes, using $\Queryalgo$.
\item  Then, it obtains the entry to exit distance $\Distance(En_j, Ex_j)$  to see if it was modified, and continues with the next iteration of $\ell+1$.
\end{compactenum}
%Note that the updates can be performed since the tree-decompositions $\Tree(G_i)$ are for the graph $(V_i,E_i')$, i.e. in which all edges between entry-exit pairs are present. 
See Algorithm~\ref{algo:rmsdist} for a formal description.

\smallskip
\begin{algorithm}%[H]
\small
\SetAlgoNoLine
\DontPrintSemicolon
%\setstretch{1.05}
\caption{$\RMSDist$}\label{algo:rmsdist}
\KwIn{A set of control flow graphs $\mathcal{G}=\{G_i\}_{1\leq i\leq k}$, stack height $h$}
%\KwOut{A local distance map $\LD_{\Bag}$ for each bag $\Bag\in V_T$}
\BlankLine
\ForEach{$G_i\in \mathcal{G}$}{
Construct the tree-decomposition $\Tree(G_i)$\\
Call $\Preprocessalgo$ on $\Tree(G_i)$\label{line:call_preprocess}\\
}
$\mathsf{distances}\leftarrow [\text{Call } \Queryalgo \text{ on } (En_i,Ex_i) \text{ of } G_i]_{1\leq i\leq k}$\label{line:call_query}\\
$\mathsf{modified}\leftarrow \{1,\dots, k\}$\\
\For{$\ell \leftarrow 0$ \KwTo $h-1$}{\label{line:loop}
$\mathsf{modified}'\leftarrow \emptyset$\\
\ForEach{$i\in \mathsf{modified}$}{
\ForEach{$G_j$ that contains boxes $b_{j_1},\dots, b_{j_l}$ s.t. $Y_j(b_{j_x})=i$}{
Call $\Updatealgo$ on $G_j$ for the weight change $\Weight((en, b_{j_l}), (ex, b_{j_x}))\leftarrow\mathsf{distances}[i]$\label{line:call_update}\\
Call $\Queryalgo$  on  $(En_j,Ex_j)$\\
\uIf{$\Distance(En_j, Ex_j)\neq \mathsf{distances}[j]$}{
$\mathsf{modified}'\leftarrow \mathsf{modified}'\cup \{j\}$\\
$\mathsf{distances}[j]\leftarrow \Distance(En_j, Ex_j)$\\
}
}
}
%\uIf{$\mathsf{modified}'=\emptyset$}{
%\textbf{break}\\
%}
$\mathsf{modified}\leftarrow \mathsf{modified}'$
}
\end{algorithm}

\smallskip\noindent{\bf Correctness and logarithmic pair query time.}
The algorithm $\RMSDist$ is described so that a proof by induction is straightforward for correctness. Initially, running the algorithm $\Preprocessalgo$ from Section~\ref{sec:dynamic} on each of the graphs $G_i$ allows queries for the distances $\Distance(u,v,0)$ for all pairs of nodes $(u,v)$, since no method call can be made. Also, the induction follows directly since for every CSM $A_i$, updating the distance from call nodes $(en,b)$ to the corresponding return nodes $(ex,b)$ of every box $b$ that corresponds to a CSM $A_j$ whose distance $\Distance(En_j,Ex_j)$ was changed in the last iteration $\ell$, ensures that the distance $\Distance(u,v,\ell+1)$ of every pair of nodes $u,v$ in $A_i$ is computed correctly. This is also true for the special pair of nodes $En_i$, $Ex_i$, which feeds the next iteration of $\RMSDist$. Finally, 
$\RMSDist$ requires $O(\sum_{i=1}^k(n_i\cdot \log n_i))$ time to construct a balanced tree decomposition (Theorem~\ref{thm:seminice}), $O(n)$ time to preprocess all $G_i$ initially, and $O(\sum_{i=1}^k(b_i\cdot \log n_i))$ to update all $G_i$ for one iteration of the loop of Line~\ref{line:call} (from Theorem~\ref{thm:dynamization}).
Hence, $\RMSDist$ uses $O(\sum_{i=1}^k (n_i\cdot \log n_i+h\cdot b_i\cdot \log n_i))$ preprocessing semiring operations. Finally, it is easy to verify that all preprocessing is done in $O(n)$ space. 

After the last iteration of algorithm $\RMSDist$, we have a datastructure $\Datastructure$ that occupies $O(n)$ space and answers distance queries $\Distance(u,v,h)$ in $O(\log n_i)$ time, with $u,v\in V_i$,  by calling $\Queryalgo$ from Theorem~\ref{sec:dynamic} for the distance $\Distance(u,v)$ in $G_i$. 

\smallskip
\begin{example}\label{ex:rsm}
We now present a small example of how $\RMSDist$ is executed on the RSM of Figure~\ref{fig:rsm} for the case of reachability.
In this case, for any pair of nodes $(u,v)$, we have $\Distance(u,v)=\True$ iff $u$ reaches $v$.
Table~\ref{tab:example}\protect\subref{subfig:preprocess_update} illustrates how the local distance maps $\LD_{\Bag_x}$
look for each bag $\Bag_x$ of each of the CSMs of the two methods dot\_vector and dot\_matrix.
Each column represents the local distance map of the
corresponding bag $\Bag_x$, and an entry $(u,v)$ means that $\LD_{\Bag_x}(u,v)=\True$ (i.e., $u$ reaches $v$).
For brevity, in the table we hide self loops (i.e., entries of the form $(u,u)$) although they are stored by the algorithms.
Initially, the stack height $\ell=0$, and $\Preprocessalgo$ is called for each graph (line~\ref{line:call_preprocess}). The new reachability relations discovered by $\Mergealgo$
are shown in bold. Note that at this point we have $\Weight(4,5)=\False$ in method dot\_matrix, as we do not know whether the call to
method dot\_vector actually returns.
Afterwards, $\Queryalgo$ is called to discover the distance $\Distance(1,6)$ in method dot\_vector (line~\ref{line:call_query}). Table~\ref{tab:example}\protect\subref{subfig:query}
shows the sequence in which $\Queryalgo$ examines the bags of the tree decomposition, and the distances $\delta_1$, $\delta_6$ and $\delta$ it maintains.
When $\Bag_2$ is examined, $\delta=\True$ and hence at the end $\Queryalgo$ returns $\delta=\True$.
Finally, since $\Queryalgo$ returns $\delta=\True$, the weight $\Weight(4,5)$ between the call-return pair of nodes $(4,5)$ in method dot\_matrix 
is set to $\True$. An execution of $\Updatealgo$ (line~\ref{line:call_update}) with this update on the corresponding tree decomposition (Table~\ref{tab:example}\protect\subref{subfig:preprocess_update} for $\ell=1$) updates the entries $(4,5)$ and
$(4,3)$ in $\LD_{\Bag_5}$ of method dot\_matrix  (shown in bold).
From this point, any same-context distance query can be answered in logarithmic time in the size of its CSM by further calls to $\Queryalgo$.

\begin{table*}%
  \centering
  \subfloat[][]{
  \renewcommand{\arraystretch}{1.1}
{\footnotesize
\makebox[\textwidth][c]{%
\begin{tabular}{| c |||  c | c | c | c | c | c || c | c | c | c | c | c | c | c | c |}
\hline
& \multicolumn{6}{c||}{\textbf{\small dot\_vector }} & \multicolumn{8}{c|}{\textbf{\small dot\_matrix }}\\
\hline
\hline
$\ell/\LD_{\Bag_x}$ & $\Bag_1$ & $\Bag_2$ & $\Bag_3$ & $\Bag_4$ & $\Bag_5$ & $\Bag_6$ & $\Bag_1$ & $\Bag_2$ & $\Bag_3$ & $\Bag_4$ & $\Bag_5$ & $\Bag_6$ & $\Bag_7$ & $\Bag_8$\\
\hline
\multirow{2}{*}{$\ell=0$}&$-$& $(1,2)$ & $(2,3)$ & $(2,3)$ & $(2,5)$ & $(5,6)$ &$-$& $(1,2)$ & $(2,3)$ & $(3,4)$ & $(3,4)$ & $(2,6)$ & $(2,7)$ & $(7,8)$ \\
&& &  &$(3,4)$ &   &  & &  &   &   & $(5,3)$ & $(3,6)$ &&\\
(Preprocess)&&&& $(4,2)$ &&&&&&&& $(6,2)$ &&\\
&&&& $\mathbf{(2,4)}$ &&&&&&&& $\mathbf{(3,2)}$ &&\\
\hline
\multirow{2}{*}{$\ell=1$} &$-$& $(1,2)$ & $(2,3)$ & $(2,3)$ & $(2,5)$ & $(5,6)$ &$-$& $(1,2)$ & $(2,3)$ & $(3,4)$ & $(3,4)$ & $(2,6)$ & $(2,7)$ & $(7,8)$ \\
&& &  &$(3,4)$ &   &  & &  &   &   & $(5,3)$ & $(3,6)$ &&\\
(Update)&&&& $(4,2)$ &&&&&&& $\mathbf{(4,5)}$ & $(6,2)$ &&\\
&&&& $(2,4)$ &&&&&&& $\mathbf{(4,3)}$ & $(3,2)$ &&\\
\hline
\end{tabular}\label{subfig:preprocess_update}
}
}
  }
  \\
  \subfloat[][]{\footnotesize
  \makebox[\textwidth][c]{%
  \renewcommand{\arraystretch}{1.1}
  \begin{tabular}{| c || c | c | c | c |}
\hline
&\multicolumn{4}{c|}{\textbf{\small dot\_vector }}\\
\hline
\hline
 &$\Bag_6$ & $\Bag_5$ & $\Bag_2$ & $\Bag_1$\\
\hline
\multirow{1}{*}{Query} &$\delta_6=\{5, 6\}$ & $\delta_6=\{2,5\}$ & $\delta_6=\{1,2\}$ & $\delta_6=\{1\}$\\
\multirow{2}{*}{$\Distance(1,6)$} &$-$ & $-$ &$\delta_1=\{1,2\}$ & $\delta_1=\{1\}$ \\
&$-$ & $-$ & $\delta=\True$ & $\delta=\True$ \\
\hline
\end{tabular}\label{subfig:query}
 }
 }
\caption{Illustration of $\RMSDist$ on the tree decompositions of methods dot\_vector and dot\_matrix from Figure~\ref{fig:rsm}.
Table~~\protect\subref{subfig:preprocess_update} shows the local distance maps for each bag and stack height $\ell=0,1$. Table~\protect\subref{subfig:query}
shows how the distance query $\Distance(1,6)$ in method dot\_vector is handled.
}
 \label{tab:example}
\end{table*}

\end{example}

\smallskip\noindent{\bf Linear single-source query time.}
In order to handle single-source queries, some additional preprocessing is required. 
The basic idea is to use $\RMSDist$ to process the graphs $G_i$, and then use additional preprocessing  on each $G_i$
by applying existing algorithms for graphs with constant treewidth.
For graphs with constant treewidth, an extension of Lemma~7 from~\cite{Chaudhuri95} 
allows us to precompute the distance $\Distance(u,v)$ for every pair of nodes $u,v\in V_i$ 
that appear in the same bag of $\Tree(G_i)$. The computation
required is similar to $\Preprocessalgo$, with the difference that this time $\Tree(G_i)$ is traversed top-down instead of bottom-up. Additionally,
for each examined bag $\Bag$, a Floyd-Warshall algorithm is run in the graph $G_i$ induced by $B$, and all pairs of distances are updated. It follows from Lemma~7 of~\cite{Chaudhuri95} that for constant treewidth, this step requires $O(n_i)$ time and space.

After all distances $\Distance(u,v)$ have been computed for each $\Bag$, it is straightforward to answer single-source queries from some node $u$ in linear time. The algorithm simply maintains a map $A:V_i\rightarrow \Sigma$, and initially $A(v)=\Distance(u,v)$ for all $v\in \Bag_u$, and $A(v)=\Zero$ otherwise. Then, it traverses $\Tree(G_i)$ in a BFS manner starting at $\Bag_u$, and for every encountered bag $\Bag$ and $v\in \Bag$, if $A(v)=\Zero$, it sets $A(v)=\bigoplus_{z\in \Bag}\bigotimes(A(z),\Distance(z,v))$. 
For constant treewidth, this results in a constant number of semiring operations per bag, and hence $O(n_i)$ time in total.

\smallskip\noindent{\bf Constant pair query time.}
After $\RMSDist$ has returned, it is possible to further preprocess the graphs $G_i$ to reduce the pair query time to constant, while increasing the space by a factor of $\log n_i$. For constant treewidth, this can be obtained by adapting Theorem~10 from~\cite{Chaudhuri95} to our setting, which in turn is based on a rather complicated algorithmic technique of~\cite{Alon87}. 
We present a more intuitive, simpler and implementable approach that has a dynamic programming nature. 
In Section~\ref{sec:results} we present some experimental results obtained by this approach.

Recall that the extra preprocessing for answering single-source queries in linear time consists in computing $\Distance(u,v)$ for every pair of nodes $u,v$ that appear in the same bag, at no overhead. To handle pair queries in constant time, we further traverse each $\Tree(G_i)$ one last time, bottom-up, and for each node $u$ we store maps $F_u,T_u:V_i^{\Bag_u}\rightarrow \Sigma$, where $V_i^{\Bag_u}$ is the subset of $V_i$ of nodes that appear in $\Bag_u$ and its descendants in $\Tree(G_i)$. The maps are such that $F_u(v)=\Distance(u,v)$ and $T_u=\Distance(v,u)$. Hence, $F_u$ stores the distances from $u$ to nodes in $V_i^{\Bag_u}$, and $T_u$ stores the distances from nodes in $V_i^{\Bag_u}$ to $u$. The maps are computed in a dynamic programming fashion, as follows:
\begin{compactenum}
\item  Initially, the maps $F_u$ and $T_u$ are constructed for all $u$ that appear in a bag $\Bag$ which is a leaf of $\Tree(G_i)$. The information required has already been computed as part of the preprocessing for answering single-source queries. Then, $\Tree(G_i)$ is traversed up, level by level.
\item When examining a bag $\Bag$ such that the computation has been performed for all its children, for every node $u\in \Bag$ and $v\in V_i^{\Bag}$, we set $F_u(v)=\bigoplus_{z\in \Bag}\bigotimes\{\Distance(u,z), F_z(v)\}$, and similarly for $T_u=\bigoplus_{z\in \Bag}\bigotimes\{\Distance(z,u), T_z(v)\}$.
\end{compactenum}
An application of Lemma~\ref{lem:tree_paths} inductively on the levels processed by the algorithm can be used to show that when a bag $\Bag$ is processed,
for every node $u\in \Bag$ and $v\in V_i^{\Bag}$, we have $T_u(v)=\bigoplus_{P:v\rightsquigarrow u}\otimes(P)$ and $F_u(v)=\bigoplus_{P:u\rightsquigarrow v}\otimes(P)$. Finally, there are $O(n_i)$ semiring operations done at each level of $\Tree(G_i)$, and since there are $O(\log n_i)$ levels, $O(n_i\cdot \log n_i)$ operations are required in total. Hence, the space used is also $O(n_i\cdot \log n_i)$. We furthermore preprocess $\Tree(G_i)$ in linear time and space to answer LCA queries in constant time (note that since $\Tree(G_i)$ is balanced, this is standard). To answer a pair query $u,v$, it suffices to first obtain the LCA $\Bag$ of $\Bag_u$ and $\Bag_v$, and it follows from Lemma~\ref{lem:tree_paths} that $\Distance(u,v)=\bigoplus_{z\in \Bag}\bigotimes\{T_z(u), F_z(v)\}$, which requires a constant number of semiring operations.

We conclude the results of this section with the following theorem. Afterwards, we obtain the results 
for the special cases of the IFDS/IDE framework, reachability and shortest path.

\smallskip
\begin{theorem}\label{thm:rmsdist}
Fix the following input: (i)~a constant treewidth RSM $A=\set{A_1,A_2,\dots, A_k}$, where $A_i$ consists of $n_i$ nodes and $b_i$ boxes; (ii)~a partially complete semiring $(\Sigma, \oplus, \otimes, \Zero, \One)$; and (iii)~a maximum stack height $h$. $\RMSDist$ uses $O(\sum_{i=1}^k (n_i\cdot \log n_i+h\cdot b_i\cdot \log n_i))$ preprocessing semiring operations and

\begin{compactenum}
\item Using $O(n)$ space it correctly answers same-context algebraic pair queries in $O(\log n_i)$, and same-context algebraic single-source queries in $O(n_i)$ semiring operations.
%\item The algorithm $\RMSDist$ correctly solves algebraic path distance queries  and uses $O(\sum_{i=1}^k (n_i\cdot \log n_i+h\cdot b_i\cdot \log n_i))$ preprocessing time and semiring operations and solves algebraic pair queries in $O(\alpha(n_i))$ time, where $A_i$ is the CSM that contains the nodes of the query.
\item Using $O(\sum_{i=1}^k(n_i\cdot \log n_i))$ space, it correctly answers same-context algebraic pair queries in $O(1)$ semiring operations.
\end{compactenum}
\end{theorem}

\smallskip\noindent{\bf IFDS/IDE framework.}
In the special case where the algebraic path problem belongs to the IFDS/IDE framework, we have a meet-composition semiring $(F, \sqcap, \circ, \emptyset,  I)$, where $F$ is a set of distributive flow functions $2^D\rightarrow 2^D$, $D$ is a set of data facts, $\sqcap$ is the meet operator (either union or intersection), $\circ$ is the flow function composition operator, and $I$ is the identity flow function. For a fair comparison, the $\circ$ semiring operation does not induce a unit time cost, but instead a cost of $O(|D|)$ per data fact (as functions are represented as bipartite graphs~\cite{Reps95}). Because the set $D$ is finite, and the meet operator is either union or intersection, it follows that the image of every data fact will be updated at most $|D|$ times. 
% the weight of every call to return edge $((en,b),(ex,b))$ of every $G_i$ that corresponds to CSM $A_i$ will be updated at most $|D|$ times for each data fact in $D$,
Then, line~\ref{line:loop} of $\RMSDist$ needs to change so that instead of $h$ iterations, the body of the loop is carried up to a fixpoint. 
The amortized cost per $G_i$ is then $ b_i\cdot \log n_i\cdot |D|^3$ (as there are $|D|$ data facts), and we have the following corollary
(also see Table~\ref{tab:IDFS-comparison}).

\smallskip
\begin{corollary}[IFDS/IDE]\label{cor:ifds}
Fix the following input a (i)~constant treewidth RSM $A=\set{A_1,A_2,\dots, A_k}$, where $A_i$ consists of $n_i$ nodes and $b_i$ boxes; and (ii)~a meet-composition semiring $(F, \sqcap, \circ, \emptyset,  I)$ where $F$ is a set of distributive flow functions $D\rightarrow D$, $\circ$ is the flow function composition operator and $\sqcap$ is the meet operator.

\begin{compactenum}
\item Algorithm $\RMSDist$ uses $O(\sum_{i=1}^k (n_i\cdot |D|^2 +b_i\cdot \log n_i\cdot |D|^3 + n_i\cdot \log n_i))$ preprocessing time, $O(n\cdot |D|^2)$ space, and correctly answers same-context algebraic pair queries in $O(\log n_i\cdot |D|^2)$ time, and same-context algebraic single-source queries in $O(n_i\cdot |D|^2)$ time.
\item Algorithm $\RMSDist$ uses $O(\sum_{i=1}^k (n_i\cdot \log n_i\cdot |D|^2+b_i\cdot \log n_i\cdot |D|^3))$ preprocessing time, $O(|D|^2\cdot \sum_{i=1}^k(n_i\cdot \log n_i))$ space, and correctly answers same-context algebraic pair queries in $O(|D|^2)$ time, and same-context algebraic single-source queries in $O(n_i\cdot |D|^2)$ time.
\end{compactenum}
\end{corollary}

\smallskip\noindent{\bf Reachability}.
The special case of reachability is obtained by setting $|D|=1$ in Corollary~\ref{cor:ifds}.

\smallskip\noindent{\bf Shortest paths.}
The shortest path problem can be formulated on the tropical semiring $(\Reals_{\geq 0}\cup \{\infty\}, \min, +, \infty,  0)$.
We consider that both semiring operators cost unit time (i.e., the weights occurring in the computation fit in a constant number of 
machine words). Because we consider non-negative weights, it follows that the distance between any pair of nodes is realized
by a path that traverses every entry node at most once. Hence, we set $h=k$ in Theorem~\ref{thm:rmsdist}, and obtain the following corollary for shortest paths
(also see Table~\ref{tab:shortest_path-comparison}).

\smallskip
\begin{corollary}[Shortest paths]\label{cor:shortest_paths}
Fix the following input a (i)~constant treewidth RSM $A=\set{A_1,A_2,\dots, A_k}$, where $A_i$ consists of $n_i$ nodes and $b_i$ boxes; (ii)~a tropical semiring $(\Reals_{\geq 0}\cup \{\infty\}, \min, +, \infty,  0)$. $\RMSDist$ uses $O(\sum_{i=1}^k (n_i\cdot \log n_i +k\cdot b_i\cdot \log n_i))$ preprocessing time and:

\begin{compactenum}
\item  Using $O(n)$ space, it correctly answers same-context shortest path pair queries in $O(\log n_i)$, and same-context shortest path single-source queries in $O(n_i)$ time.
\item Using  $O(\sum_{i=1}^k(n_i\cdot \log n_i))$ space, it correctly answers same-context shortest path pair queries in $O(1)$ time.
\end{compactenum}
\end{corollary}

\smallskip\noindent{\bf Interprocedural witness paths.}
As in the case of simple graphs from Section~\ref{sec:dynamic}, we can retrieve a witness path
for any distance $\Distance(u,v,h)$ that is realized by acyclic interprocedural paths $P:(u,\emptyset)\rightsquigarrow (v,\emptyset)$,
without affecting the stated complexities. The process is straightforward. Let $A_i$ contain the 
pair of nodes $u,v$ on which the query is asked. Initially, we obtain the witness
intraprocedural path $P':u\rightsquigarrow v$, as described in Section~\ref{sec:dynamic}. 
Then, we proceed recursively to obtain a witness path $P_j$
between the entry $En_j$ and exit $Ex_j$ nodes of every CSM $A_j$ such that $P'$ contains
an edge between a call node $(en, b)$ and a return node $(ex, b)$ with $Y_i(B)=j$.
That is, we reconstruct a witness path for every call to a CSM whose weight has been summarized
locally in $A_i$.
%%It is easy to see that 
This process constructs an interprocedural witness path $P:u\rightsquigarrow v$
such that $\otimes(P)=\Distance(u,v)$ in $O(|P|)$ time.

\section{Experimental Results}\label{sec:results}
\begin{table*}[ht]
\renewcommand{\arraystretch}{1.1}
{\scriptsize
\makebox[\textwidth][c]{%
\begin{tabular}{| c ||  c | c ||| c | c || c | c | c | c ||| c | c || c | c | c | c | }
\hline
%\multicolumn{4}{|c||}{}& \multicolumn{3}{c||}{\textbf{Our Preprocessing}} & \multicolumn{2}{c||}{\textbf{No Preprocessing}} & \multicolumn{1}{c|}{\textbf{Complete Preprocessing}}\\
\multicolumn{3}{|c|||}{\textbf{\small Benchmarks}}& \multicolumn{6}{c|||}{\textbf{\small Interprocedural Reachability}} & \multicolumn{6}{c|}{\textbf{\small Intraprocedural Shortest path}}\\
\hline
\multicolumn{3}{|c|||}{} & \multicolumn{2}{c||}{\textbf{Preprocessing}} & \multicolumn{4}{c|||}{\textbf{Query}} & \multicolumn{2}{c||}{\textbf{Preprocessing}} & \multicolumn{4}{c|}{\textbf{Query}}\\
\hline
\multicolumn{3}{|c|||}{} & \multicolumn{2}{c||}{\textbf{}} & \multicolumn{2}{c|}{\textbf{Single}} & \multicolumn{2}{c|||}{\textbf{Pair}} & \multicolumn{2}{c||}{\textbf{}} & \multicolumn{2}{c|}{\textbf{Single}} & \multicolumn{2}{c|}{\textbf{Pair}}\\
\hline
& $n$ & $t$ & Our & Complete & Our & No Prepr. & Our & No Prepr. & Our & Complete & Our & No Prepr. & Our & No Prepr.\\
\hline
\hline
antlr & 698 & 1.0 & 76316 & 136145 & 15.3 & 166.3 & 0.15 & 14.34 & 221578 & 1.13$\cdot 10^7$&251& 24576 & 0.36 & 24576  \\
\hline
bloat & 696 & 2.3 & 27597 & 54335 & 3.9 & 72.5 & 0.10 & 14.34 & 87950 & 1.15$\cdot 10^7$&257& 25239 & 0.37 & 25239  \\
\hline
chart & 1159 & 1.5 & 22191 & 90709 & 2.3 & 80.9 & 0.13 & 22.32 & 125468 & 1.24$\cdot 10^8$&398& 88856 & 0.39 & 88856  \\
\hline
eclipse & 656 & 1.6 &  37010 & 138905 & 6.7 & 239.1 & 0.19 & 15.76 &  152293 & 1.07$\cdot 10^7$&533& 23639 & 0.46 & 23639  \\ 
\hline
fop & 1209 & 1.7  &  30189 & 91795 & 2.9 & 60.6 & 0.12 & 43.0 &   153728 & 3.94$\cdot 10^8$&1926& 113689 & 2.71 & 113689  \\
\hline
hsqldb & 698 & 1.0  & 55668 & 180333 & 13.0 & 219.0 & 0.14 & 13.89 & 215063 & 1.23$\cdot 10^7$&236& 24322 & 0.36 & 24322  \\
\hline
jython & 748 & 1.5 & 43609 & 68687 & 7.2 & 85.7 & 0.11 & 12.84 & 159085 & 1.42$\cdot 10^7$&386& 29958 & 0.32 & 29958  \\
\hline
luindex & 885 & 1.3  &  36015 & 142005 & 5.6 & 202.7 & 0.16 & 26.44 &  163108 & 2.97$\cdot 10^7$ &258& 51192 & 0.37 & 51192  \\
\hline
lusearch & 885 & 1.3 & 51375 & 189251 & 12.8 & 211.4 & 0.13 & 26.01 &  219015 & 2.90$\cdot 10^7$&254& 50719 & 0.34 & 50719  \\
\hline
pmd & 644 & 1.4  &  31483 & 52527 & 2.5 & 83.9 & 0.13 & 12.5 &  140974 & 9.14$\cdot 10^6$ &327& 22572 & 0.37 & 22572  \\
\hline
xalan & 698 & 1.0 & 57734 & 138420 & 8.0 & 235.0 & 0.19 & 14.28 &  186695 & 1.10$\cdot 10^7$&380& 24141 & 0.43 & 24141  \\
\hline
Jflex & 1091 & 1.6  & 51431 & 91742 & 3.1 & 50.8 & 0.11 & 20.46 & 154818 & 1.24$\cdot 10^8$&231& 83093 & 0.36 & 83093  \\
\hline
muffin & 1022 & 1.7  & 29905 & 66708 & 2.6 & 52.7 & 0.10 & 18.57 &  125938 & 1.02$\cdot 10^8$&265& 80878 & 0.38 & 80878  \\
\hline
javac & 711 & 1.8 & 32981 & 59793 & 4.8 & 75.2 & 0.11 & 11.86 &  117390 & 1.31$\cdot 10^7$&370& 26180 & 0.34 & 26180  \\
\hline
polyglot & 698 & 1.0 & 68643 & 150799 & 12.2 & 184.5 & 0.14 & 14.14 & 228758 & 1.15$\cdot 10^7$&244& 24400 &  0.35 & 24400  \\
\hline
\end{tabular}
}
}
\caption{Average statistics gathered from our experiments on the DaCapo benchmark suit. Times are in microseconds.}\label{tab:results1}
\end{table*}

\noindent{\bf Set up.}
We have implemented our algorithms for linear-time single-source and constant-time 
pair queries presented in Section~\ref{sec:rsm} and have tested them on graphs obtained from the 
DaCapo benchmark suit~\cite{Blackburn06} that contains several, real-world Java applications.
Every benchmark is represented as a RSM that consists of several CSMs, 
and each CSM corresponds to the control flow graph of a method of the benchmark.
We have used the Soot framework~\cite{Soot} for obtaining the control flow graphs,
where every node of the graph corresponds to one Jimple statement of Soot,
and the tool of~\cite{Dijk06} to obtain their tree decompositions. 
Our experiments were run on a standard desktop computer with a 3.4GHz CPU, on a single thread.

\smallskip\noindent{\bf Interprocedural reachability and intraprocedural shortest path.}
In our experiments, we focus on the important special case of reachability and 
shortest path.
We consider CSMs of moderate to large size (all CSMs with at least five hundred nodes),
as for small CSMs the running times are negligible.
The first step is to execute an interprocedural reachability algorithm from the program entry
to discover all actual call to return edges $((en,b), (ex,b))$ of every CSM $A_i$
(i.e., all invocations that actually return), and then consider the control flow graphs 
$G_i$ independently. 
\begin{compactitem}
\item \emph{(Reachability).}
For every $G_i$, the complete preprocessing in the case of reachability is done by executing $n_i$ 
DFSs, one from each source node. The single-source query from $u$ is answered by executing 
one DFS from $u$, and the pair query $u,v$ is done similarly, but we stop as soon as $v$ is 
reached.
We note that this methodology correctly answers interprocedural same-context reachability 
queries.

\item \emph{(Shortest path).}
For shortest path we perform intraprocedural analysis on each $G_i$.
We assign both positive and negative weights to each edge of $G_i$ uniformly at random
from the range $[-10,10]$.
For general semiring path properties, the Bellman-Ford algorithm~\cite{Cormen01}
is a very natural one, which in the case of shortest path can handle positive and
negative weights, as long as there is no negative cycle.
To have a meaningful comparison with Bellman-Ford (as a representative of a general
semiring framework), we consider both positive and negative weights, but do not allow
negative cycles.
For complete preprocessing we run the classical Floyd-Warshall algorithm
(which computes all-pairs shortest paths and is a generalization of Bellman-Ford).
Under no preprocessing, for every single-source and pair query we run the Bellman-Ford algorithm.
\end{compactitem}

\smallskip\noindent{\bf Results.}
Our experimental results are shown in Table~\ref{tab:results1}.
\begin{compactenum}

\item The average treewidth of control flow graphs is confirmed to be very small,
and does not scale with the size of the graph. In fact, even the largest treewidth is four.

\item The preprocessing time of our algorithm is significantly less than the complete 
preprocessing, by factor of~1.5 to~4 times in case of reachability, and 
by orders of magnitude in case of shortest path.

\item In both reachability and shortest path, all queries are handled significantly faster after our preprocessing,
than no preprocessing.
We also note that for shortest path queries, Bellman-Ford answers single-source and pair queries 
in the same time, which is significantly slower than both our single-source and pair queries.
% and a fair comparison is against our single-source queries which is significantly faster.
Finally, we note that for single-source reachability queries, though we do not provide
theoretical improvement over DFS (Table~\ref{tab:IDFS-comparison}), the one-time 
preprocessing information allows for practical improvements.

\end{compactenum}

Since our work focuses on same-context queries and the IFDS/IDE framework does not have this restriction,
 a direct comparison with the IFDS/IDE framework 
would be biased in our favor.
In the experimental results for interprocedural reachability with same-context queries, 
we show that we are faster than even DFS (which is faster than IFDS/IDE).

\smallskip\noindent{\bf Description of Table~\ref{tab:results1}.}
In the table, the second (resp. third) column shows the average number of nodes (resp. 
treewidth) of CSMs of each benchmark.
The running times of preprocessing are gathered by averaging over all CSMs in each benchmark.
The running times of querying are gathered by averaging over all possible single-source and pair queries
in each CSM, and then averaging over all CSMs in each benchmark.

\section{Conclusions}\label{sec:conclusions}
In this work we considered constant treewidth RSMs since control flow graphs 
of most programs have constant treewidth. 
We presented algorithms to handle multiple same-context algebraic path queries,
where the weights belong to a partially complete semiring.
Our algorithms have small additional one-time preprocessing, 
but answer subsequent queries significantly faster than no preprocessing
both in terms of theoretical bounds as well as in practice, even for basic 
problems such as reachability and shortest path. 
While in this work we focused on RSMs with unique entries and exits,
an interesting theoretical question is 
to extend our results to RSMs with multiple entries and exists.

\smallskip\noindent{\bf Acknowledgements.}
We thank anonymous reviewers for helpful comments to 
improve the presentation of the paper.

\bibliographystyle{unsrt}
\bibliography{bibliography}

\begin{thebibliography}{10}

\bibitem{Reps95}
Thomas Reps, Susan Horwitz, and Mooly Sagiv.
\newblock Precise interprocedural dataflow analysis via graph reachability.
\newblock In {\em POPL}, New York, NY, USA, 1995. ACM.

\bibitem{Sagiv96}
Mooly Sagiv, Thomas Reps, and Susan Horwitz.
\newblock Precise interprocedural dataflow analysis with applications to
  constant propagation.
\newblock {\em Theor. Comput. Sci.}, 1996.

\bibitem{Callahan86}
David Callahan, Keith~D. Cooper, Ken Kennedy, and Linda Torczon.
\newblock Interprocedural constant propagation.
\newblock In {\em CC}. ACM, 1986.

\bibitem{Grove93}
Dan Grove and Linda Torczon.
\newblock Interprocedural constant propagation: A study of jump function
  implementation.
\newblock In {\em PLDI}. ACM, 1993.

\bibitem{Land91}
William Landi and Barbara~G. Ryder.
\newblock Pointer-induced aliasing: A problem classification.
\newblock In {\em POPL}. ACM, 1991.

\bibitem{Knoop96}
Jens Knoop, Bernhard Steffen, and J\"{u}rgen Vollmer.
\newblock Parallelism for free: Efficient and optimal bitvector analyses for
  parallel programs.
\newblock {\em ACM Trans. Program. Lang. Syst.}, 1996.

\bibitem{Cousot77}
P.~Cousot and R~Cousot.
\newblock Static determination of dynamic properties of recursive procedures.
\newblock In E.J. Neuhold, editor, {\em IFIP Conf. on Formal Description of
  Programming Concepts}, 1977.

\bibitem{Giegerich81}
Robert Giegerich, Ulrich M\"{o}ncke, and Reinhard Wilhelm.
\newblock Invariance of approximate semantics with respect to program
  transformations.
\newblock In {\em 3rd Conference of the European Co-operation in Informatics
  (ECI)}, 1981.

\bibitem{Knoop92}
Jens Knoop and Bernhard Steffen.
\newblock The interprocedural coincidence theorem.
\newblock In {\em CC}, 1992.

\bibitem{Naeem08}
Nomair~A. Naeem and Ondrej Lhot{\'a}k.
\newblock Typestate-like analysis of multiple interacting objects.
\newblock In {\em OOPSLA}, 2008.

\bibitem{Zhang14}
Xin Zhang, Ravi Mangal, Mayur Naik, and Hongseok Yang.
\newblock Hybrid top-down and bottom-up interprocedural analysis.
\newblock In {\em PLDI}, 2014.

\bibitem{Chatterjee15}
Krishnendu Chatterjee, Andreas Pavlogiannis, and Yaron Velner.
\newblock Quantitative interprocedural analysis.
\newblock In {\em POPL}, 2015.

\bibitem{ABEGRY05}
R.~Alur, M.~Benedikt, K.~Etessami, P.~Godefroid, T.~W. Reps, and M.~Yannakakis.
\newblock Analysis of recursive state machines.
\newblock {\em ACM Trans. Program. Lang. Syst.}, 2005.

\bibitem{Robertson84}
Neil Robertson and P.D Seymour.
\newblock Graph minors. iii. planar tree-width.
\newblock {\em Journal of Combinatorial Theory, Series B}, 1984.

\bibitem{Halin76}
Rudolf Halin.
\newblock S-functions for graphs.
\newblock {\em Journal of Geometry}, 1976.

\bibitem{Thorup98}
Mikkel Thorup.
\newblock {All Structured Programs Have Small Tree Width and Good Register
  Allocation}.
\newblock {\em Information and Computation}, 1998.

\bibitem{Gustedt02}
Jens Gustedt, OleA. Mæhle, and JanArne Telle.
\newblock The treewidth of java programs.
\newblock In {\em Algorithm Engineering and Experiments}, LNCS. Springer, 2002.

\bibitem{Reps05}
Thomas Reps, Stefan Schwoon, Somesh Jha, and David Melski.
\newblock Weighted pushdown systems and their application to interprocedural
  dataflow analysis.
\newblock {\em Sci. Comput. Program.}, 2005.

\bibitem{Reps07}
Thomas Reps, Akash Lal, and Nick Kidd.
\newblock Program analysis using weighted pushdown systems.
\newblock In {\em FSTTCS 2007: Foundations of Software Technology and
  Theoretical Computer Science}, LNCS. 2007.

\bibitem{Chaudhuri95}
Shiva Chaudhuri and Christos~D. Zaroliagis.
\newblock {Shortest Paths in Digraphs of Small Treewidth. Part I: Sequential
  Algorithms}.
\newblock {\em Algorithmica}, 1995.

\bibitem{Cormen01}
T.H. Cormen, C.E. Leiserson, R.L. Rivest, and C.~Stein.
\newblock {\em {Introduction To Algorithms}}.
\newblock MIT Press, 2001.

\bibitem{Fischer71}
Michael~J. Fischer and Albert~R. Meyer.
\newblock {Boolean Matrix Multiplication and Transitive Closure}.
\newblock In {\em SWAT (FOCS)}. IEEE Computer Society, 1971.

\bibitem{Dijk06}
Thomas van Dijk, Jan-Pieter van~den Heuvel, and Wouter Slob.
\newblock Computing treewidth with libtw.
\newblock Technical report, University of Utrecht, 2006.

\bibitem{Horwitz95}
Susan Horwitz, Thomas Reps, and Mooly Sagiv.
\newblock Demand interprocedural dataflow analysis.
\newblock {\em SIGSOFT Softw. Eng. Notes}, 1995.

\bibitem{ATM06}
R.~Alur, S.~{La Torre}, and P.~Madhusudan.
\newblock Modular strategies for recursive game graphs.
\newblock {\em Theor. Comput. Sci.}, 2006.

\bibitem{CV12}
Krishnendu Chatterjee and Yaron Velner.
\newblock Mean-payoff pushdown games.
\newblock In {\em LICS}, 2012.

\bibitem{Chaudhuri08}
Swarat Chaudhuri.
\newblock Subcubic algorithms for recursive state machines.
\newblock In {\em POPL}, New York, NY, USA, 2008. ACM.

\bibitem{Arnborg89}
Stefan Arnborg and Andrzej Proskurowski.
\newblock {Linear time algorithms for NP-hard problems restricted to partial
  k-trees }.
\newblock {\em Discrete Appl Math}, 1989.

\bibitem{Bern1987216}
M.W Bern, E.L Lawler, and A.L Wong.
\newblock Linear-time computation of optimal subgraphs of decomposable graphs.
\newblock {\em J Algorithm}, 1987.

\bibitem{Bodlaender88}
Hans~L. Bodlaender.
\newblock Dynamic programming on graphs with bounded treewidth.
\newblock In {\em ICALP}, LNCS. Springer, 1988.

\bibitem{Bodlaender93}
Hans~L. Bodlaender.
\newblock A tourist guide through treewidth.
\newblock {\em Acta Cybern.}, 1993.

\bibitem{Bodlaender05}
HansL. Bodlaender.
\newblock Discovering treewidth.
\newblock In {\em SOFSEM 2005: Theory and Practice of Computer Science}, volume
  3381 of {\em LNCS}. Springer, 2005.

\bibitem{Courcelle91}
Brouno Courcelle.
\newblock Graph rewriting: An algebraic and logic approach.
\newblock In {\em Handbook of Theoretical Computer Science (Vol. B)}. MIT
  Press, Cambridge, MA, USA, 1990.

\bibitem{Elberfeld10}
M.~Elberfeld, A.~Jakoby, and T.~Tantau.
\newblock Logspace versions of the theorems of bodlaender and courcelle.
\newblock In {\em FOCS}, 2010.

\bibitem{Bodlaender94}
HansL. Bodlaender.
\newblock Dynamic algorithms for graphs with treewidth 2.
\newblock In {\em Graph-Theoretic Concepts in Computer Science}, LNCS.
  Springer, 1994.

\bibitem{Hagerup00}
Torben Hagerup.
\newblock {Dynamic algorithms for graphs of bounded treewidth}.
\newblock {\em {Algorithmica}}, 2000.

\bibitem{L13}
J.~Lacki.
\newblock Improved deterministic algorithms for decremental reachability and
  strongly connected components.
\newblock {\em ACM Transactions on Algorithms}, 2013.

\bibitem{CL13}
K.~Chatterjee and J.~Lacki.
\newblock Faster algorithms for {Markov} decision processes with low treewidth.
\newblock In {\em CAV}, 2013.

\bibitem{Obdrzalek03}
Jan Obdrz{\'a}lek.
\newblock Fast mu-calculus model checking when tree-width is bounded.
\newblock In {\em CAV}, 2003.

\bibitem{Reed92}
Bruce~A. Reed.
\newblock Finding approximate separators and computing tree width quickly.
\newblock In {\em STOC}, 1992.

\bibitem{Kloks94}
Ton Kloks.
\newblock {\em {Treewidth, Computations and Approximations}}.
\newblock LNCS. Springer, 1994.

\bibitem{Sridharan05}
Manu Sridharan, Denis Gopan, Lexin Shan, and Rastislav Bod\'{\i}k.
\newblock Demand-driven points-to analysis for java.
\newblock In {\em OOPSLA}, 2005.

\bibitem{Alon87}
Noga Alon and Baruch Schieber.
\newblock {Optimal preprocessing for answering on-line product queries}.
\newblock Technical report, Tel Aviv University, 1987.

\bibitem{Blackburn06}
Stephen M. et~al. Blackburn.
\newblock The dacapo benchmarks: Java benchmarking development and analysis.
\newblock In {\em OOPSLA}, 2006.

\bibitem{Soot}
Raja Vall{\'e}e-Rai, Phong Co, Etienne Gagnon, Laurie Hendren, Patrick Lam, and
  Vijay Sundaresan.
\newblock Soot - a java bytecode optimization framework.
\newblock In {\em CASCON '99}. IBM Press, 1999.

\end{thebibliography}

\end{document}